 \documentclass{aa}  

\usepackage{txfonts}
\usepackage{mathastext}
\usepackage{natbib}
\usepackage{graphicx}
\bibpunct{(}{)}{;}{a}{}{,} 
\usepackage{longtable}
\usepackage{xspace}
 \newcommand\micron{\mbox{$\mu$m}\xspace}

\begin{document}

\title{Rotational properties of the binary and non-binary populations in the Trans-Neptunian belt \thanks{Table 2 is only available in electronic form in the Center of astronomical
Data of Strasbourg (CDS) via anonymous ftp to cdsarc.u-strasbg.fr (130.79.128.5).
} }


\author{A. Thirouin \inst{1,2} \thanks{This work has been done partly during a stay at the NASA-Goddard Space Flight Center supported by a MEC sub-project (BES-2009-014574). }
\and K.S. Noll \inst{2} \and J.L. Ortiz \inst{1} \and N. Morales \inst{1}  }

\offprints{Audrey Thirouin \email{thirouin@iaa.es}}

\institute{$^1$ Instituto de Astrof\'{\i}sica  de Andaluc\'{\i}a, CSIC, Apt 3004, 18080, Granada, Spain. \\
$^2$ NASA Goddard Space Flight Center, Greenbelt MD 20771, United States of America.}

 
\abstract{We present results for the short-term variability of Binary Trans-Neptunian Objects (BTNOs). We performed CCD photometric observations using the 3.58~m Telescopio Nazionale Galileo (TNG), the 1.5~m Sierra Nevada Observatory (OSN) telescope, and the 1.23~m Centro Astron\'{o}mico Hispano Alem\'{a}n (CAHA) telescope at Calar Alto Observatory. We present results based on five years of observations and report the short-term variability of six BTNOs. Our sample contains three classical objects: (174567) 2003~MW$_{12}$, or Varda, (120347) 2004~SB$_{60}$, or Salacia, and 2002~VT$_{130}$; one detached disk object: (229762) 2007~UK$_{126}$; and two resonant objects: (341520) 2007~TY$_{430}$ and (38628) 2000~EB$_{173}$, or Huya. For each target, possible rotational periods and/or photometric amplitudes are reported. We also derived some physical properties from their lightcurves, such as density, primary and secondary sizes, and albedo. We compiled and analyzed a vast lightcurve database for Trans-Neptunian Objects (TNOs) including centaurs to determine the lightcurve amplitude and spin frequency distributions for the binary and non-binary populations. The mean rotational periods, from the Maxwellian fits to the frequency distributions, are 8.63$\pm$0.52~h for the entire sample, 8.37$\pm$0.58~h for the sample without the binary population, and 10.11$\pm$1.19~h for the binary population alone. Because the centaurs are collisionally more evolved, their rotational periods might not be so primordial. We computed a mean rotational period, from the Maxwellian fit, of 8.86$\pm$0.58~h for the sample without the centaur population, and of 8.64$\pm$0.67~h considering a sample without the binary and the centaur populations. According to this analysis, regular TNOs spin faster than binaries, which is compatible with the tidal interaction of the binaries. 
Finally, we examined possible formation models for several systems studied in this work and by our team in previous papers.} 


   \maketitle
%

\section{Introduction}

As of December 2013, 78 binary/multiple systems \footnote{The exact definition of binarity is the following: a system composed of two objects orbiting their common center of mass or barycenter, which lies outside either body. For example, in the Trans-Neptunian belt, the Pluto-Charon system is a true binary. For most of the binary/multiple systems in this belt we have no information about their barycenter, so the use of this term has to be considered carefully. Often, the term binary/multiple is used to refer to systems with one or more companions despite the definition mentioned here.} have been identified in the Trans-Neptunian belt. The majority of such systems have only one satellite, but two systems are known to have two companions, Haumea and 1999~TC$_{36}$ \citep{Brown2006, Benecchi2010}, while the Pluto system consists of the Pluto/Charon binary accompanied by four relatively small satellites. This means that $\sim$5$\%$ of the known Trans-Neptunian Objects (TNOs) have at least one companion. However, some estimations indicate that the proportion of these systems must be up to 20-25$\%$ \citep{Noll2008}.  

The discovery of binary/multiple systems in the Trans-Neptunian belt is subject to observational limitations. In fact, large telescopes, typically 4m class telescopes, are required to detect companion(s). The first Binary TNO (BTNO) (apart from Charon) was the near-equal-sized companion of 1998~WW$_{31}$ \citep{Veillet2002}. This detection was a serendipitous discovery using the 3.6~m Canada France Hawaii Telescope under excellent weather and seeing conditions. Several large ground-based surveys dedicated to the discovery of TNOs had some sensitivity for detecting (or not) binaries, such as the \textit{Deep Ecliptic Survey (DES)} \citep{Millis2002, Elliot2005}, the \textit{Deep Keck Search for Binary Kuiper Belt Objects} \citep{Schaller2003}, and the \textit{Canada France Ecliptic Plane Survey (CFEPS)} \citep{Jones2006}. Ground-based surveys have detected only a few binaries, and especially near-equal-sized systems with a large separation between the components. The most prolific tool for detecting such systems is the \textit{Hubble Space Telescope (HST)} which has a high rate of discoveries \citep{Noll2008}.    

A binarity orbital study supplies a vast set of information, such as mass, density, albedo, size of each of the components of the system (under some assumptions), as well as important clues about formation and evolution of the Trans-Neptunian belt \citep{Noll2008}. Some approaches can be used to complement the binarity study, such as spectroscopy \citep{Carry2011}, and photometric studies. The short-term variability allows us to retrieve rotation periods from the photometric periodicities and also provides constraints on shape (or surface heterogeneity) by means of the lightcurve amplitude. Unfortunately, fewer than 30 BTNOs have a well-determined rotational period. The sample is highly biased because the majority of BTNOs (and TNOs in general) observed are large and bright objects with high lightcurve variability and short rotational period \citep{Thirouin2010, Thirouin2012}. In this work, we increase the number of BTNOs whose short-term variability has been studied. From performing CCD photometric observations using several telescopes in Spain during the past five years, we report six BTNO short-term variability studies. From the lightcurve, we derived some physical characteristics, such as density constraints, albedo, and primary/satellite sizes. Part of this paper is dedicated to the rotational properties of the binary and non-binary populations. We study several spin frequency distributions to pinpoint evidence of tidal effects between the components of multiple systems. 

This paper is divided into six sections. In the next section, we describe the observations and the data set. Section 3 describes the reduction techniques we used to derive periods and photometric ranges. In Section 4, we summarize our main results for each target. In Section 5, we discuss our results and present a summary about binarity and non-binarity in the Trans-Neptunian belt. Finally, Section 6 is dedicated to our conclusions.   
  

\section{Observations}
\subsection{Runs and telescopes}

We present data obtained with the 3.58~m Telescopio Nazionale Galileo (TNG), with the 1.5~m Sierra Nevada Observatory (OSN) telescope, and with the 1.23~m Centro Astron\'{o}mico Hispano Alem\'{a}n (CAHA) telescope at Calar Alto Observatory between 2009 and 2012.

The TNG is located at the Roque de los Muchachos Observatory (La Palma, Canary Islands, Spain). Images were obtained using the Device Optimized for the LOw RESolution instrument (DOLORES, or LRS). This device has a camera and a spectrograph installed at the Nasmyth B telescope focus. We observed in imaging mode with the R Johnson filter and a 2$\times$2 binning mode. The camera is equipped with a 2048x2048 CCD with a pixel size of 13.5$\micron$. The field of view is 8.6$\arcmin$$\times$8.6$\arcmin$ with a 0.252$\arcsec$/pix scale (pixel scale for a 1$\times$1 binning).

The 1.5~m telescope is located at the Observatory of Sierra Nevada (OSN), at Loma de Dilar in the National Park of Sierra Nevada (Granada, Spain). Observations were carried out by means of a 2k$\times$2k CCD, with a total field of view of 7.8$\arcmin$$\times$7.8$\arcmin$. We used a 2$\times$2 binning mode, which changes the image scale to 0.46$\arcsec$/pixel.
 
The 1.23~m CAHA telescope at Calar Alto Observatory is located in the Sierra de Los Filabres (Almeria, Spain). Observations were carried out by means of a 4k$\times$4k CCD, with a total field of view of 21.5$\arcmin$$\times$21.5$\arcmin$ (in a 1$\times$1 binning) and a pixel scale of 15~$\micron$. We observed with the R Johnson filter and a 2$\times$2 binning mode.

\subsection{Observing strategy}

Exposure times were chosen by considering two main factors: i) exposure time had to be long enough to achieve a signal-to-noise ratio (S/N) sufficient to study the observed object (typically, S/N$>$20); ii) exposure time had to be short enough to avoid elongated images of the target (telescope tracked at sidereal speed) or elongated field stars (telescope tracked at the object rate of motion). We always chose to track the telescope at sidereal speed. The drift rates of TNOs are low, typically $\sim$2$\arcsec$/h, so exposure times around 400 to 700 seconds were used.

Observations at the OSN were performed without filter to maximize the S/N. As the main goal of our study is short-term variability via relative photometry, the use of unfiltered images without absolute calibration is not a problem. The R Johnson filter was used during our observations with the TNG and with the 1.23~m CAHA telescope. This filter was chosen to maximize the object S/N and to minimize the fringing that appears at longer wavelengths in these instruments.

In this work, we focused on six BTNOs: Salacia, Varda, 2002~VT$_{130}$, 2007~UK$_{126}$, 2007~TY$_{430}$, and Huya. All relevant geometric information about the observed objects at the dates of observations, and the number of images and filters used are summarized in Table~\ref{Log_Obs}. 

\begin{scriptsize}
 
\begin{table*}
\caption {\label{Log_Obs} Dates (UT-dates, format MM/DD/YYYY), heliocentric (r$_{h}$), and geocentric ($\Delta$) distances, and phase angle ($\alpha$). We also indicate the number of images (Nb. Images) obtained each night, the filter and telescope for each observational run. OSN stands for Observatory of Sierra Nevada, TNG for Telescopio Nazionale Galileo, and 1.23~m CAHA for the 1.23~m Centro Astron\'{o}mico Hispano Alem\'{a}n (CAHA) telescope at Calar Alto Observatory.
} 
 \begin{scriptsize}
\begin{tabular}{@{}cccccccc}  
\hline 
 Object  & Date & Nb. Images & r$_{h}$ [AU] & $\Delta$ [AU] & $\alpha$ [$^{\circ}$] & Filter & Telescope \\ \hline

(174567) 2003~MW$_{12}$ or Varda & 07/24/2009 & 6 & 47.812 & 47.235 & 1.01 & R &TNG \\
 & 07/25/2009 & 12 & 47.812 & 47.247 & 1.02 & R &TNG \\
 & 07/27/2009 & 7 & 47.812 & 47.260 & 1.03 & R &TNG \\
 & 07/03/2011 & 27 & 47.573 & 46.746 & 0.72 & R &TNG \\
 & 07/04/2011 & 29 & 47.573 & 46.754 & 0.73 & R &TNG \\
 & 07/28/2011 & 26 & 47.565 & 47.002 & 1.03 &Clear& OSN\\
 & 07/29/2011 & 30 & 47.564 & 47.015 & 1.04 &Clear& OSN\\
 & 07/30/2011 & 9 & 47.564 & 47.029 & 1.05 &Clear& OSN\\
 & 06/13/2012 & 8  & 47.456  &  46.519 & 0.48 &Clear& OSN\\
 & 06/14/2012 & 6  &  47.456 & 46.522  & 0.49 &Clear& OSN\\
 & 06/15/2012 & 32  &  47.455 & 46.524 & 0.50 &Clear& OSN\\
 & 07/04/2013 & 28  &  47.325  & 46.492  & 0.71 & Clear & OSN \\
 & 07/05/2013 & 7  & 47.325  & 46.500 &  0.73 & Clear & OSN \\
 & 07/06/2013 & 44  &  47.325 &  46.507 &  0.74 & Clear & OSN \\
 
(120347) 2004~SB$_{60}$ or Salacia & 06/30/2011 & 4 & 44.234 & 43.991 & 1.28 & R &TNG \\
 & 07/01/2011 & 22 & 44.235 & 43.975 & 1.28 &  R &TNG \\
 & 10/31/2011 & 42 & 44.263 & 43.617 & 0.98 &  R &TNG \\
 & 09/13/2012 & 8  &   44.335 & 43.413 & 0.53  &Clear& OSN\\
 & 09/14/2012 &  28 &  44.335  &  43.413 &  0.52  &Clear& OSN\\
 & 09/15/2012 & 34  &  44.336 & 43.413 &  0.52 &Clear& OSN\\
 & 09/16/2012 & 39  &  44.336   & 43.413  &  0.52  &Clear& OSN\\
 & 10/12/2012 & 15  &  44.342 &  43.514 & 0.73  &Clear& OSN\\
 & 10/15/2012 & 25 &  44.342 & 43.537 &  0.77   &Clear& OSN\\
 & 08/07/2013 &  27 &    44.408 &   43.705 &  0.96  &Clear& OSN\\
 & 08/08/2013 &  43 &   44.409  &  43.695 &   0.97 &Clear& OSN\\

(341520) 2007~TY$_{430}$ & 10/28/2011 & 18  & 29.041 &28.057 & 0.28 &R& TNG\\ 
 & 10/29/2011 & 19  & 29.041  &28.059 & 0.31 &R& TNG\\ 
 & 10/31/2011 & 17  & 29.040 &28.066 & 0.38 &R& TNG\\ 
 & 11/01/2011 & 36  & 29.040 &28.069 & 0.41 &R& TNG\\ 

(229762) 2007~UK$_{126}$  & 10/28/2011 & 54  & 44.515 &43.688 & 0.72 &R& TNG\\ 
 & 10/30/2011 & 42  & 44.513  &43.673 & 0.69 &R& TNG\\ 
 & 10/31/2011 & 26  & 44.512  &43.665 & 0.67 &R& TNG\\ 

(38628) 2000~EB$_{173}$ or Huya & 06/07/2010 & 21 & 28.676 &27.852 & 1.20 &Clear& OSN\\
 & 06/10/2010 & 9 & 28.676 & 27.880& 1.28 &Clear& OSN\\
 & 06/11/2010 &19&28.676  &27.890 & 1.30 &Clear& OSN\\
 & 05/25/2012 &43 & 28.578 & 27.632& 0.74 &R& 1.23~m CAHA\\
 & 05/26/2012 & 8 & 28.578 & 27.637& 0.77 &R& 1.23~m CAHA\\
 & 05/29/2012 & 39 & 28.578 &27.651 & 0.84 &R& 1.23~m CAHA\\
 & 06/12/2012 &15 & 28.576 &27.752 & 1.21 &Clear& OSN\\
 & 06/14/2012 &19 & 28.576 &27.771 & 1.26 &Clear& OSN\\
 & 06/09/2013 & 35 & 28.549  &   27.675&  1.06 &Clear& OSN\\
 & 06/10/2013 & 37  & 28.548  &27.683  &  1.08 &Clear& OSN\\
 & 06/11/2013 & 37 &  28.548 &  27.691 & 1.11  &Clear& OSN\\
 & 06/13/2013 &  17 &  28.548 &  27.707 & 1.16  &Clear& OSN\\

2002~VT$_{130}$  & 11/01/2011 & 31 & 42.926 & 42.090 & 0.72 & R &  TNG\\

\hline\hline
 \end{tabular}
\end{scriptsize}
\end{table*}
 
\end{scriptsize}


\section{Data reduction and analysis}
\subsection{Data reduction}

During all observing nights, series of biases and flat-fields were obtained to correct the images. Median bias and median flat-field frames were created for each observational night. Care was taken not to use bias and/or flat-field frames that might be affected by observational and/or acquisition problems. The median flat-fields were assembled from twilight-dithered images, and the results were inspected for possible residuals from very bright saturated stars. Flat-field exposure times were always long enough to ensure that no shutter effect was present, so that a gradient or an artifact of some sort could be present in the corrected images. Each target image was bias-subtracted and flat-fielded using the median bias and median flat-field of the corresponding observing night.   

Relative photometry using between 10 to 25 field stars was carried out by means of the Daophot routines \citep{Stetson1987}. We rejected images in which the target was affected by a cosmic-ray hit or by a nearby star. Care was taken not to introduce spurious results due to faint background stars or galaxies in the aperture. We used a common reduction software for photometry data reduction of all the images, adjusting the details of the parameters to the specificity of each data set.  

The choice of the aperture radius is crucial. The aperture has to be as small as possible to obtain the highest S/N by minimizing the sky contribution. But, on the other hand, the aperture has to be large enough to include most of the target flux. Generally, we repeated the measurement using a set of apertures with radii around the full width at half maximum.  
Various reference star sets were used to obtain the relative photometry of all the objects. Several stars had to be rejected from the analysis because they presented some variability. 
The aperture size and the reference stars used were the two main factors to consider the validity of our data reduction. The star set and aperture value that gave the lowest scatter in the photometry of the stars of similar brightness to the target and of the target was used for the final result.  
The final photometry of our targets was computed by taking the median of all the lightcurves obtained with respect to each reference star. By applying this technique, spurious results were eliminated and the dispersion of photometry was improved. 

During all observational campaigns, we studied the same field of view, and therefore the same reference stars for each observed object.
When we combined data from several observing runs, we normalized the photometry data to their average because we did not have absolute photometry that would have allowed us to link runs. By normalizing over the averages of several runs, we assume that a similar number of data points are in the upper and lower part of the curves. This may not be so if runs were only two or three nights long, which is not usually the case. We wish to emphasize that we normalized to the average of each run, not the average of each night.


\subsection{Period-detection methods and single/double peaked lightcurve}
 
The time-series photometry of each target was inspected for periodicities by means of the Lomb technique \citep{Lomb1976} as implemented in \cite{Press1992}. We also checked our results with several other time-series analysis techniques, such as the phase dispersion minimization (PDM) \citep{Stellingwerf1978}, and the CLEAN technique \citep{Foster1995}. The method developped by \cite{Harris1989} and its improvement \citep{Pravec1996} was also used (Pravec-Harris method).

Finally, to measure the full amplitude (or peak-to-peak amplitude) of short-term variability, a first- or second-order Fourier fit (depending on whether we considered a single- or double-peaked rotational periodicity) to the data was performed. 

\begin{figure}
\includegraphics[width=9.5cm, angle=0]{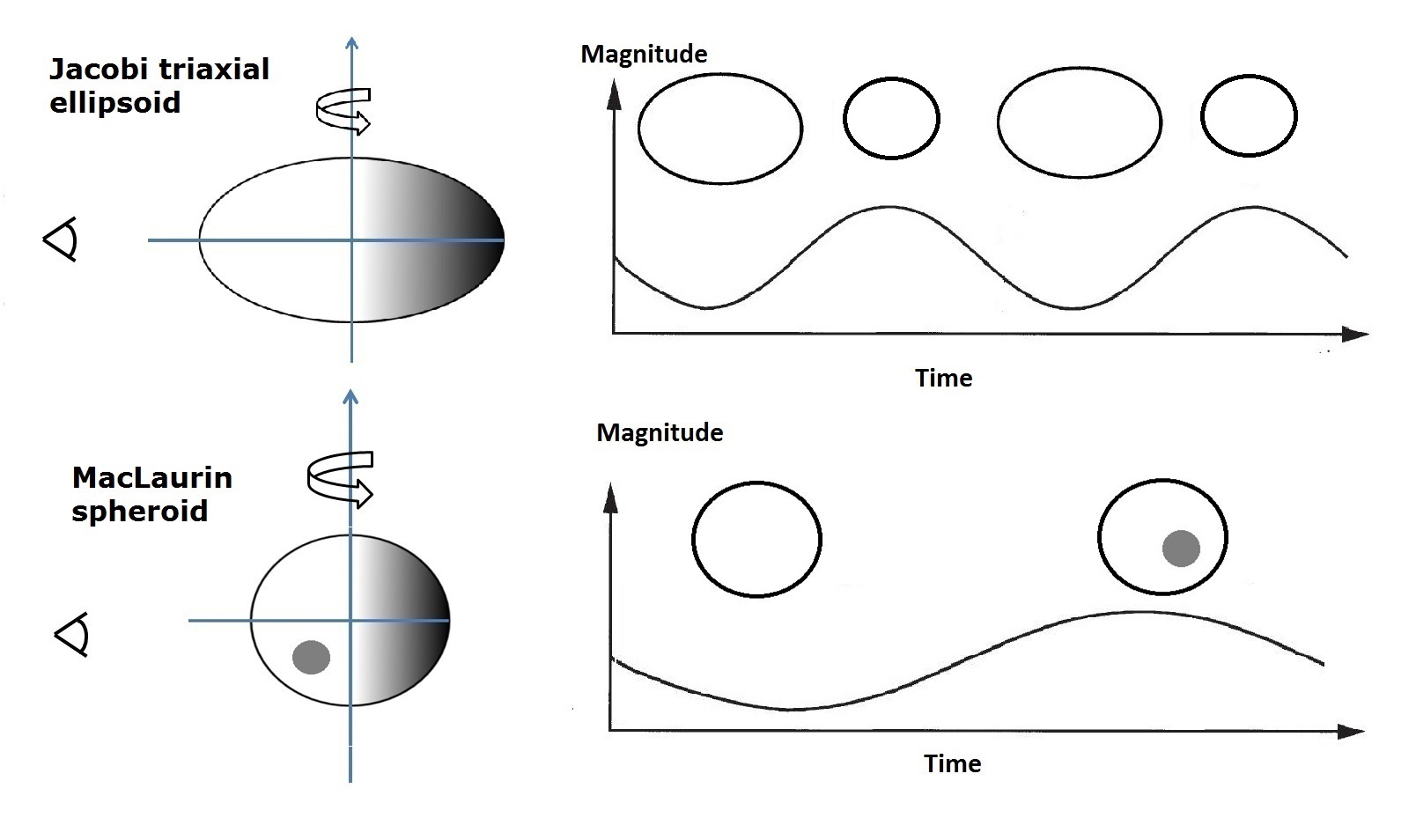}
\caption {Upper panel: assuming TNOs as triaxial ellipsoids, with axes a>b>c (rotating
around the short axis c corresponding to the lowest energy state of rotation), we obtain two brightness maxima and two minima per rotation cycle (double-peaked lightcurve) as illustrated in the curve. The maxima correspond to the largest cross-section of the object projected in the direction to the observer. Lower panel: if the TNO is a MacLaurin spheroid with an albedo variation on its surface, we obtained one maximum and one minimum per rotation cycle (single-peaked lightcurve). Note that the ordinates are magnitudes, not brightness.  }
\label{fig:SingleDouble}
\end{figure}

One important point is to distinguish between single- and double-peaked lightcurves. Except for a pole-on view of an object, in which no rotational variability can be observed, the observer will detect rotational variability for the rest of configurations of the spin axis. Assuming a triaxial ellipsoid (also known as Jacobi ellipsoid), we have to expect a lightcurve with two maxima and two minima, corresponding to a full
rotation (Figure~\ref{fig:SingleDouble}, upper panel). However, if the object is spherical or oblate (also known as MacLaurin spheroid) without any albedo variation on its surface, we have to expect a flat lightcurve. If this spheroid presents albedo variation on its surface, we have to expect a lightcurve with one maximum and one minimum (i.e., a single-peaked lightcurve). In real cases, we have combinations of shape effects and albedo variations. In many cases, when the lightcurve amplitude is (very) small, it is very difficult or impossible to distinguish whether a lightcurve is single- or double-peaked. Then we have to find a criterion to distinguish between shape and albedo effects.   

In \cite{Thirouin2010} and \cite{Duffard2009}, we proposed a threshold at 0.15~mag to distinguish whether a lightcurve is due to the albedo or due to the object shape. We know that this is a simplification because there may be elongated objects whose rotational variability is smaller than 0.15~mag simply because their rotation axes are viewed close to pole-on from Earth. However, these probably are only a small fraction of all the objects with variability below 0.15~mag because statistically only very few objects have spin axes near the pole-on orientation. \citet{Thirouin2013_phd} determined the best lightcurve amplitude limit for distinguishing between shape- and albedo-dominated lighcurves (i.e., distinguishing between single- and double-peaked lighcurves). We tested three lightcurve amplitude ($\Delta$m) limits: i) a threshold at $\Delta$m=0.10 mag, ii) at $\Delta$m=0.15 mag, and iii) at $\Delta$m=0.20 mag, to distinguish between single- and double-peaked lightcurves. For example, using the first threshold, we consider that lightcurves with an amplitude smaller than or equal to 0.10 mag are single-peaked (i.e., equivalent to assuming that the lightcurve variation is due to albedo markings), and lighcurves with an amplitude higher than 0.10 mag are double-peaked (i.e., equivalent to assuming that the lightcurve variation is due to the elongated shape of the body). This has a profound effect on the final spin-period distribution. One of the main results of this study is that the best Maxwellian fit (with a significance level of 99$\%$) is obtained for a distribution compiled assuming a threshold of 0.15~mag. Thus, it seems that this value is a good measurement of the typical variability caused by albedo. On the other hand, the high-amplitude lightcurves of large TNOs, which we can clearly attribute to a triaxial-ellipsoid shape, can indicate that the typical magnitude of hemispheric albedo changes if we compare the two maxima or two minima in the double-peaked lightcurves. This is because these objects are so large that they are in hydrostatic equilibrium, and therefore the differences in their maxima and minima can only come from albedo changes on their surfaces. The brightness differences for 2003~VS$_{2}$ and Haumea are around 0.04~mag \citep{Thirouin2010}, for 2007~TY$_{430}$ the difference is around 0.05~mag (this work), whereas for Varuna the greatest difference is 0.1~mag \citep{Thirouin2010}. This means that the hemispherically averaged albedo has variations of about 4 to 10$\%$. We expect that the variability induced by surface marks is about 0.1~mag. In the asteroid case, albedo variations are usually responsible for lightcurve amplitudes between 0.1~mag and 0.2~mag \citep{Magnusson1991, Lupishko1983, Degewij1979}. Here, we chose an intermediate value of 0.15~mag as the most reliable threshold above which we can be nearly confident that the variations are caused by shape effects. This value has been used by several investigators as the transition from low to medium variability \citep{Sheppard2008}. We must point out that distinguishing between shape and/or albedo contribution(s) in a lightcurve is not trivial at all, so a criterion must be used. 


In Table~\ref{Summary_photo}, we indicate the preferred photometric period, which is the rotational period obtained from our data reduction. The preferred rotational period, which is the rotational period assuming our criterion, is also listed. 
For example, for 2007~TY$_{430}$, the data analysis suggested a photometric single-peaked rotational period, but given that it shows an amplitude larger than 0.15~mag, the amplitude variation is probably caused by the shape of the object, therefore the double-peaked rotational period as the true rotational period for this object is preferred. 

\subsection{Satellite contribution to the lightcurve}

Because we are study binaries (no eclipsing nor contact binaries), one has to keep in mind a possible contribution of the satellite in the photometry (and in turn in the lightcurve). Neither of the two components of the system are resolved in our data, and the magnitude of the pair is measured.
 
For a wide system with a long orbital period and large separation between the two components, the satellite contribution to the lightcurve is negligible. For a very faint satellite, its lightcurve contribution is also negligible. But systems with a short orbital period (typically a few days) and a small separation between the two components requires more attention. Various systems in our sample have an orbital period of about five days. Our observational runs are, generally, one week long, which means that an entire (or nearly entire) orbital period is covered. Depending on the geometry of the system, mutual events between the primary and the satellite can be observed. With only a few fully known orbits, predicting mutual events is difficult. Only two systems (except for contact binaries) have undergone or are undergoing mutual events: Pluto-Charon \citep{Binzel1997}, and Sila-Nunam \citep{Grundy2012, Benecchi2014}. Five of the six systems studied in this paper do not have a fully known orbit, therefore we checked each observational night for possible mutual events between the primary and the satellite. No mutual event was detected during our short-term variability runs.


\section{Photometric results}

In this section, we present the short-term variability results that are summarized in Table~\ref{Summary_photo}. 
The lightcurves and Lomb periodograms for all objects are provided as online material. We only present an example of a Lomb periodogram and a lightcurve in Fig.~\ref{fig:Lomb_TY430} and Fig.~\ref{fig:LC_TY430}. We plotted all lightcurves over two cycles (rotational phase from 0 to 2) for a better visualization of the cyclical variation. Times for zero phase are reported in Table~\ref{Summary_photo}, without light-time correction and at the beginning of the integration. For each lightcurve, a first- or second-order Fourier series is used to fit the photometric data. Error bars for the measurements are not shown in the plots for clarity, but one-sigma error bars on the relative magnitudes are reported in the supplementary material (see Table~\ref{allphoto}). We must point out that when we combined several observing runs obtained at different epochs, light-time correction of the data is required. 


\begin{scriptsize}
\begin{table}

\caption{\label{allphoto} Time-series photometry of all the objects is provided at the Center of astronomical Data of Strasbourg (CDS). We present our photometric results: the name of the object, and for each image we specify the Julian date (JD, not corrected for light time), the relative magnitude (mag in magnitudes) and the 1-$\sigma$ error associated (err. in magnitude), the filter (fil.) used during observational runs, the phase angle ($\alpha$, in degree), topocentric (r$_{h}$) and heliocentric ($\Delta$) distances (both distances expressed in AU).  } 
\begin{tiny}
\begin{tabular}{@{}lccccccc} 
\hline
Object  & JD  & mag.   & Err.  &   Fil. & $\alpha$   & r$_\mathrm{h}$  & $\Delta$    \\
  &  [2450000+] & [mag] &  [mag]&     &  [$^{\circ}$] &  [AU] &  [AU]   \\
\hline
\hline
 
174567 &	 	&	 	& 	& 	&	 &	 	&	  	\\
  &	5037.43042	&	-0.023	&	0.007	&	R	&	1.01	&	47.235	&	47.812	\\
&	5037.46122	&	-0.036	&	0.008	&	R	&	1.01	&	47.235	&	47.812	\\
&	5037.46882	&	-0.011	&	0.007	&	R	&	1.01	&	47.235	&	47.812	\\
\hline
\hline

\end{tabular}
\end{tiny}
\end{table}
 
\end{scriptsize}



\subsection{The system (174567) 2003~MW$_{12}$ (or Varda) and Ilmar\"{e}}

Using \textit{Hubble Space Telescope} images obtained in April 26$^{th}$ 2009, the discovery of a satellite (Ilmar\"{e}) was reported in 2011 \citep{Grundy2011DPS}. Ilmar\"{e} is faint, with an apparent magnitude difference \footnote{The apparent magnitude difference or component magnitude difference ($\Delta$$_{mag}$) is the difference of magnitude between the satellite magnitude and the primary magnitude.} of $\sim$1.45~mag in the F606W band \footnote{\textit{Hubble Space Telescope} filter}. 
  
\begin{figure}
\includegraphics[width=9cm, angle=180]{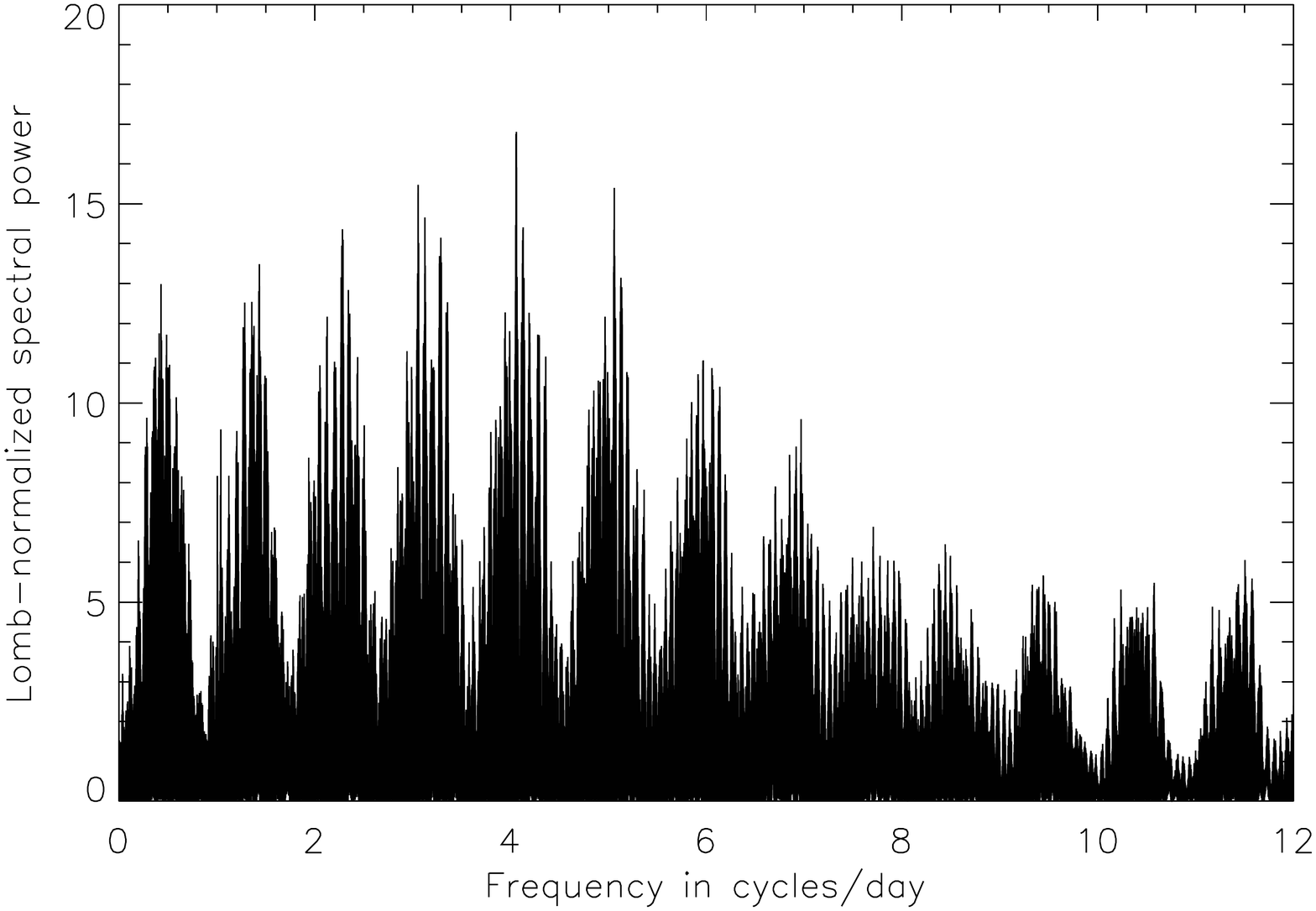}
\caption {\textit{Lomb-normalized spectral power versus frequency in cycles/day for Varda}: the Lomb periodogram of our data sets shows that one highest peak is located at 5.91~h (4.06~cycles/day) and the two largest aliases are at 7.87~h (3.04~cycles/day) and at 4.76~h (5.04~cycles/day).}
\label{fig:Lomb_MW12}
\end{figure}

\begin{figure}
\includegraphics[width=9cm, angle=0]{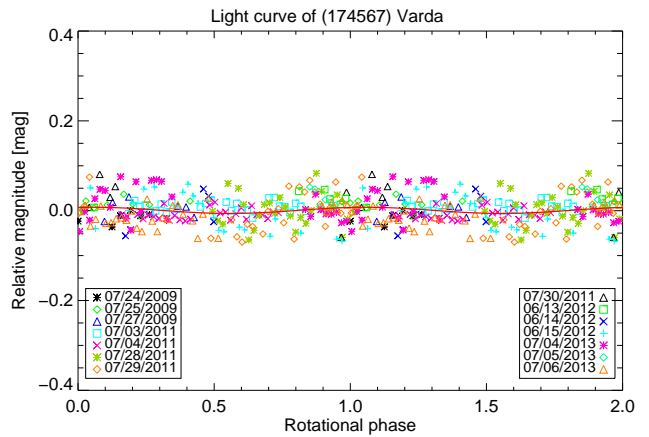}
\caption{\textit{Varda lightcurve}: rotational phase curve for Varda obtained by using a spin period of 5.91~h. The continuous line is a Fourier series fit of the photometric data. Different symbols correspond to different dates.}
\label{fig:LC_MW12}
\end{figure}

Varda was observed in July 2009 and 2011 with the TNG, in June 2012 and in July 2013 at the OSN. The Lomb periodogram (Figure~\ref{fig:Lomb_MW12}) shows several peaks. The highest peak with a significance level of 95$\%$ is located at 5.91~h (4.06~cycles/day) and the two 24-h aliases \footnote{The 24-h alias because of daylight is P$^{-1}_{alias}$=1.0027k$\pm$P$^{-1}_{real}$, where (1.0027)$^{-1}$ is the length of the sidereal day, k is an integer, and P$_{real}$ is the true rotational period of the observed object.} are at 7.87~h (3.04~cycles/day) and at 4.76~h (5.04~cycles/day). The first alias seems to have a higher confidence level than the second. All techniques (PDM, CLEAN, and Pravec-Harris method) inferred a rotational period of 5.91~h or 7.87~h. A 5.91~h rotational period is favored with a higher confidence level and therefore appears to be the best option. In Figure~\ref{fig:LC_MW12} we plot the corresponding single-peaked lightcurve with an amplitude of 0.02$\pm$0.01~mag. We must point out that for very low amplitude objects, it is difficult to estimate a secure rotational period. In fact, small variations in the night-to-night photometry can transmit more power to/from a 24~h-alias from/to the main peak. This means that we cannot completely discard the 7.87~h or the 4.76~h single-peaked rotational period.  

%
%
%
\subsection{The system (120347) 2004~SB$_{60}$ (or Salacia) and Actaea}
Using \textit{Hubble Space Telescope} images, \citet{ActaeaMPC} reported the discovery of a satellite (named Actaea) with an apparent magnitude difference of about 2.36~mag in the F606W band.    
 
\begin{figure}
\includegraphics[width=7cm, angle=90]{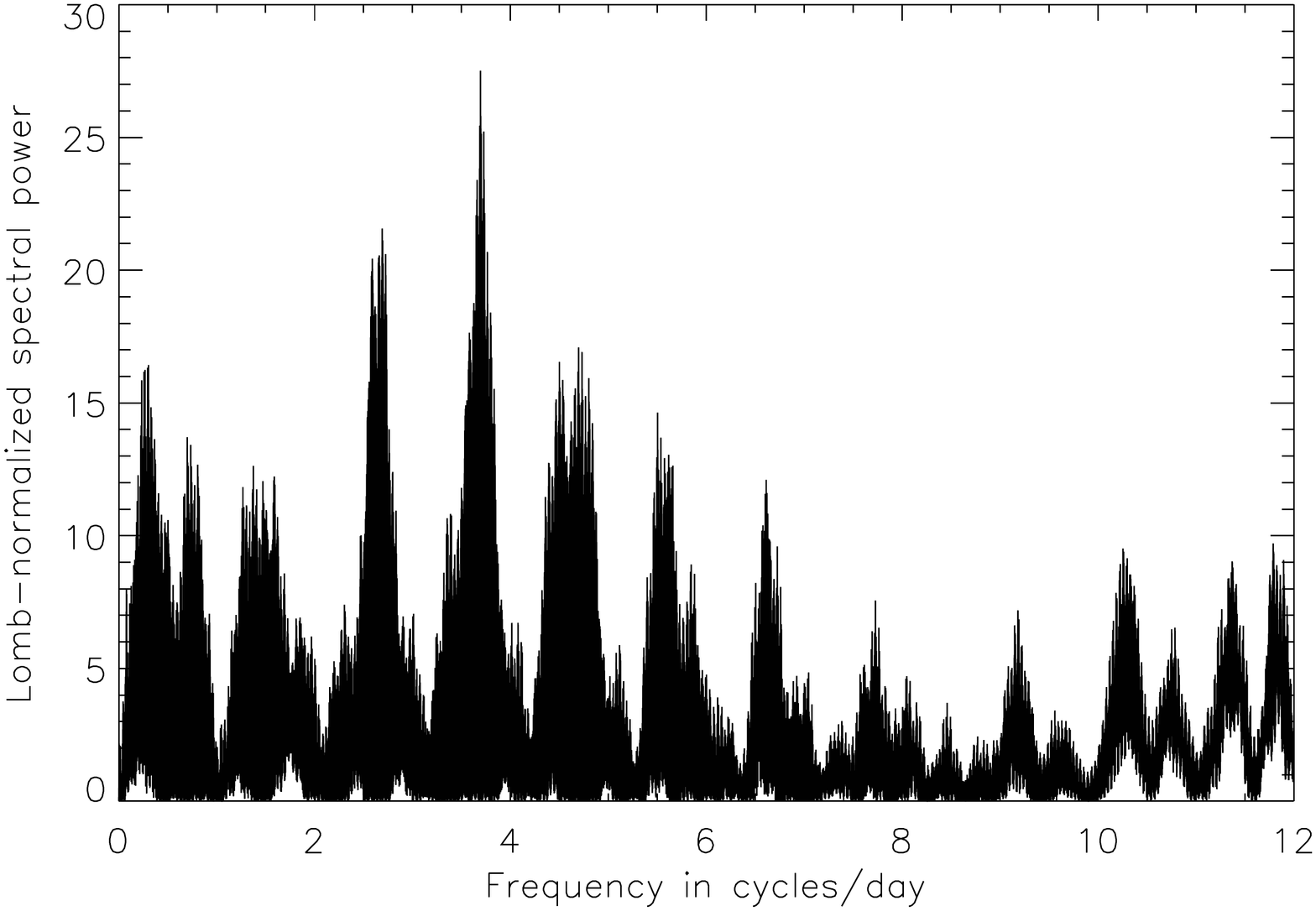}
\caption{\textit{Lomb-normalized spectral power versus frequency in cycles/day for Salacia}: the Lomb periodogram shows one main peak located at 3.69~cycles/day (6.5~h) and one alias at 2.71~cycles/day (8.86~h). }
\label{fig:Lomb_Salacia}
\end{figure}

\begin{figure}
\includegraphics[width=9cm, angle=0]{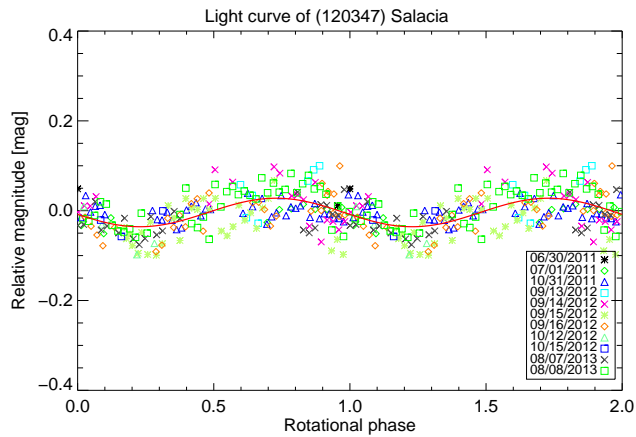}
\caption{\textit{Salacia lightcurve}: rotational phase curve for Salacia obtained by using a spin period of 6.5~h. The continuous line is a Fourier series fit of the photometric data. Different symbols correspond to different dates.}
\label{fig:LC_Salacia}
\end{figure}

Salacia was observed in July and October 2011 with the 3.58~m TNG, in September and October 2012, and in August 2013 at the OSN. The Lomb periodogram (Figure~\ref{fig:Lomb_Salacia}) shows one main peak with the highest spectral power (significance level of 99$\%$ ) located at 3.69~cycles/day (6.5~h) and the one alias with a lower spectral power located at 2.71~cycles/day (8.86~h). All techniques inferred the spin period of 6.5~h with the highest spectral power. In Figure~\ref{fig:LC_Salacia}, the corresponding single-peaked lightcurve with a peak-to-peak amplitude of 0.06$\pm$0.02~mag is plotted.

\subsection{The system 2002~VT$_{130}$}

\label{sec:VT130}

Using \textit{Hubble Space Telescope} images, \citet{Noll2009} announced the discovery of a satellite with an apparent magnitude difference of 0.44~mag in the F606W band.  
2002~VT$_{130}$  was observed during only one night in 2011 with the TNG. From about 4~h of observations, we report a 0.21~mag amplitude variation. We searched for a rotational periodicity but, unfortunately, with only few observational hours, we are not able to propose a reliable rotational-period estimation. To our knowledge, there is no bibliographic reference to compare our results with.


%
\subsection{The system (229762) 2007~UK$_{126}$}

\citet{Grundy2011DPS} reported the discovery of a companion with a magnitude difference of 3.79~mag in the F606W band. 

\begin{figure}
\includegraphics[width=9cm, angle=180]{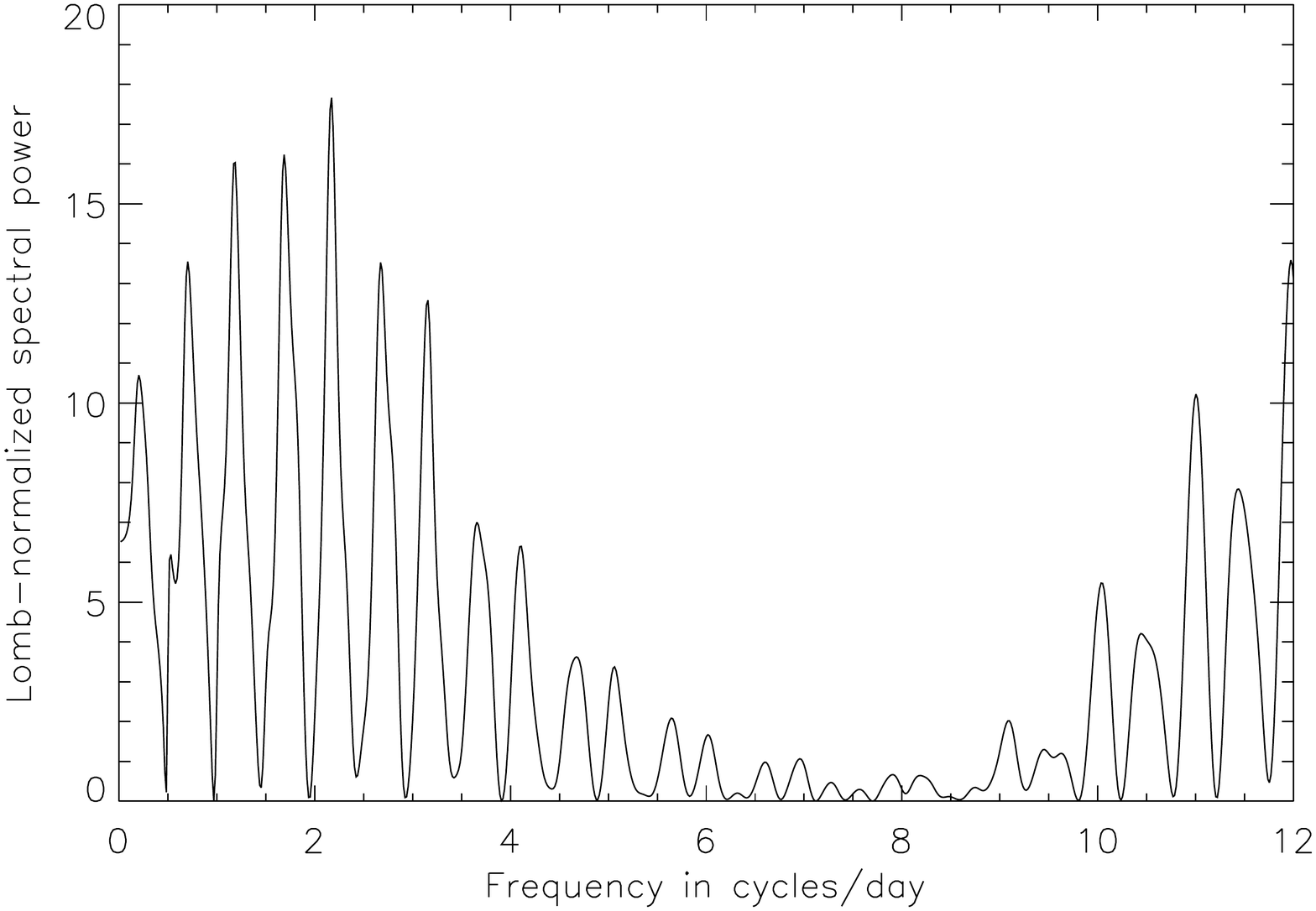}
\caption{\textit{Lomb-normalized spectral power versus frequency in cycles/day for 2007~UK$_{126}$}: the Lomb periodogram shows one peak with the highest spectral power located at 11.04~h (2.17~cycles/day), and several aliases located at 14.30~h (1.68~cycles/day) and at 20.25~h (1.19~cycles/day)}.
\label{fig:Lomb_UK126}
\end{figure}

\begin{figure}
\includegraphics[width=9cm, angle=0]{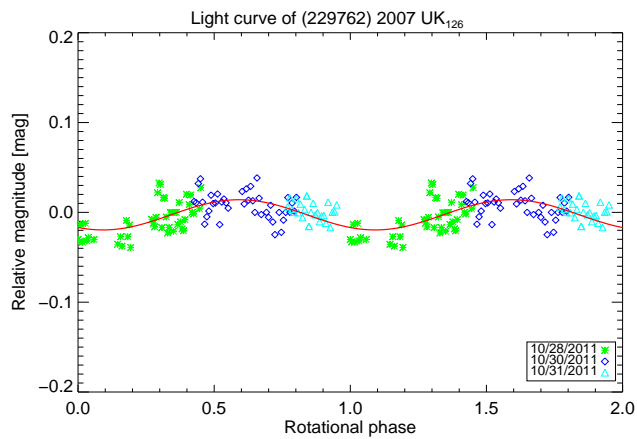}
\caption{\textit{2007~UK$_{126}$ lightcurve}: rotational phase curve for 2007~UK$_{126}$ obtained by using a spin period of 11.05~h. The continuous line is a Fourier series fit of the photometric data. Different symbols correspond to different dates. }
\label{fig:LC_UK126}
\end{figure}

2007~UK$_{126}$ was observed on October 2011 with the TNG. We report three observational nights with a time base (time coverage between the first and the last image of the night) of about 4~h, $\sim$4~h, and $\sim$2~h, respectively. 
The Lomb periodogram (Figure~\ref{fig:Lomb_UK126}) shows one main peak with a significance level of 95$\%$ located at 11.05~h (2.17~cycles/day), and several peaks located at 14.30~h (1.68~cycles/day) and at 20.25~h (1.19~cycles/day). All techniques confirmed these peaks, with a slight preference for the peak at 11.05~h. 2007~UK$_{126}$ presents a nearly flat lightcurve with a photometric variation of 0.03$\pm$0.01~mag (Figure~\ref{fig:LC_UK126}). We must point out that the presence of numerous aliases with significant spectral power in the Lomb periodogram complicatse the study and we are not able to propose a secure rotational period based on our data. We can only conclude that this object probably has a long rotational period ($>$8~h). 

\subsection{The system (341520) 2007~TY$_{430}$}

From data acquired with the 8.1~m Gemini telescope (Hawaii, USA), \cite{2007TY430MPCBinary} confirmed the presence of a satellite. The apparent magnitude difference is very low, $\sim$0.1~mag (mean value) for this equal-sized system \citep{Sheppard2012}.  

\begin{figure}
\includegraphics[width=9cm, angle=180]{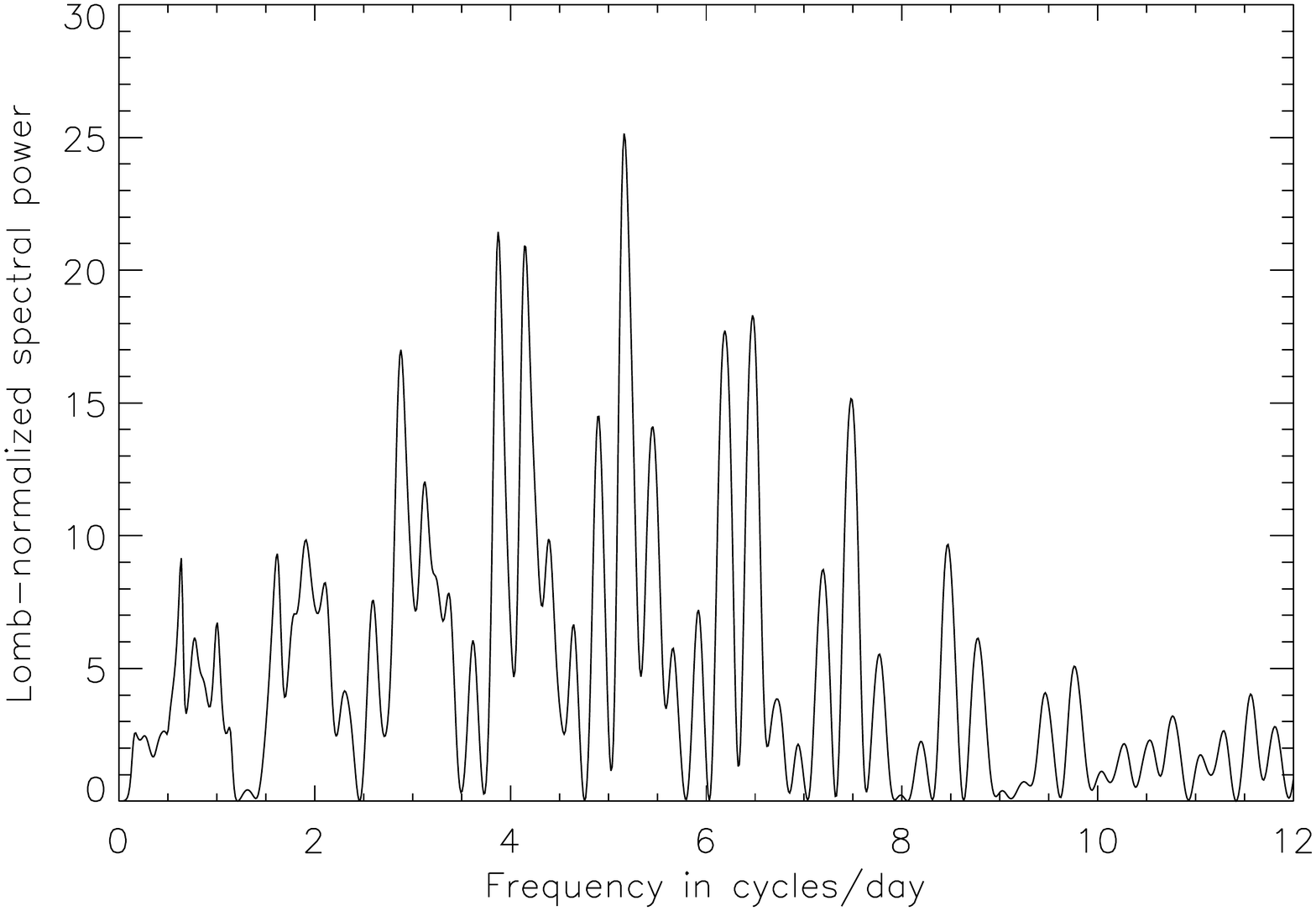}
\caption{\textit{Lomb-normalized spectral power versus frequency in cycles/day for 2007~TY$_{430}$}: the Lomb periodogram shows one main peak located at 4.64~h (5.17~cycles/day).}
\label{fig:Lomb_TY430}
\end{figure}

\begin{figure}
\includegraphics[width=9cm, angle=0]{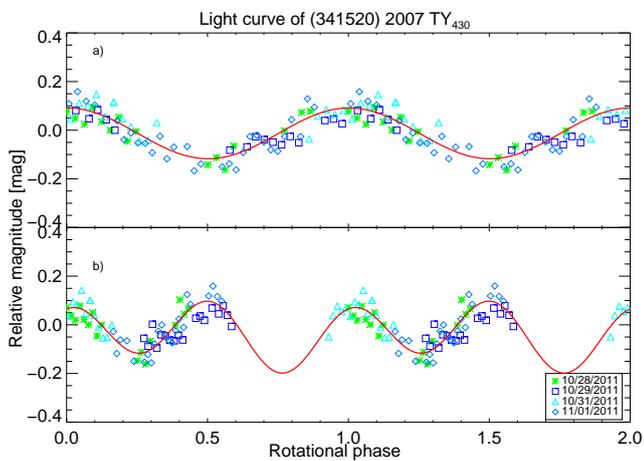}
\caption{\textit{2007~TY$_{430}$ lightcurve}: rotational phase curves for 2007~TY$_{430}$ obtained by using a spin period of 4.64~h (plot a)) and using a spin period of 9.28~h (plot b)). The continuous lines are a Fourier series fits of the photometric data. Different symbols correspond to different dates. The same legend is used in both plots. }
\label{fig:LC_TY430}
\end{figure}

2007~TY$_{430}$ was observed during about 14~h, in four nights with the TNG in 2011. The Lomb periodogram (Figure~\ref{fig:Lomb_TY430}) shows one clear peak with a high confidence level (99$\%$), located at 4.64~h (5.17~cycles/day). All techniques confirmed this periodicity. In Figure~\ref{fig:LC_TY430}, we present the single-peaked lightcurve with a rotational period of 4.64~h and an amplitude of 0.20$\pm$0.03~mag. The lightcurve amplitude is large, therefore, according to our definition (see section 3.2), we have to consider the double-peaked rotational period, 9.28~h, as true rotational period. The double-peaked lightcurve has an amplitude peak-to-peak of 0.24$\pm$0.05~mag (Figure~\ref{fig:LC_TY430}). We must point out that the first maximum is slightly taller than the second ($\sim$0.05~mag). This difference confirms the irregular shape of 2007~TY$_{430}$ and clearly favors the double-peaked lightcurve. However, we do not know if the variation comes from the primary or the satellite because both objects are apparently of very similar size which prevents us from distinguishing their contributions to the lightcurve.

\subsection{The system (38628) 2000~EB$_{173}$ (or Huya) }

2000~EB$_{173}$ (hereinafter Huya) is a resonant object and belongs to the plutino subcategory (i.e., it belongs to the 3:2 mean motion resonance with Neptune). From data acquired with the \textit{Hubble Space Telescope}, \cite{Noll2012} confirmed the discovery of a satellite with an apparent magnitude difference around 1.4~mag in the F606W band.    

\begin{figure}
\includegraphics[width=7cm, angle=90]{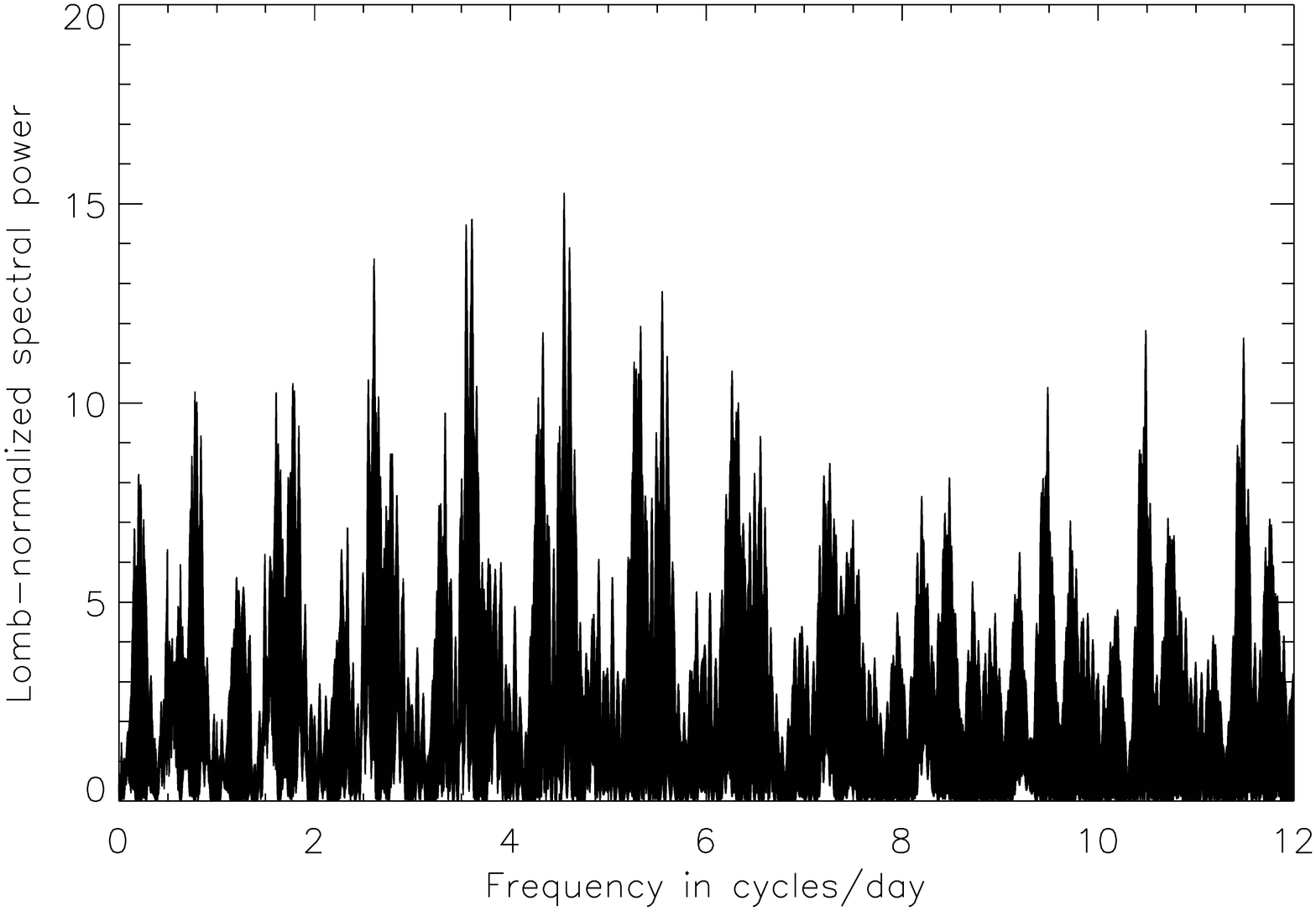}
\caption {\textit{Lomb-normalized spectral power versus frequency in cycles/day for Huya}: the Lomb periodogram shows one main peak located at 5.28~h (4.55~cycles/day), and three aliases located at 6.63~h (3.62~cycles/day), at 4.31~h (5.57~cycles/day), and at 9.15~h (2.62~cycles/day).}
\label{fig:Lomb_Huya}
\end{figure}
\begin{figure}
\includegraphics[width=9cm, angle=0]{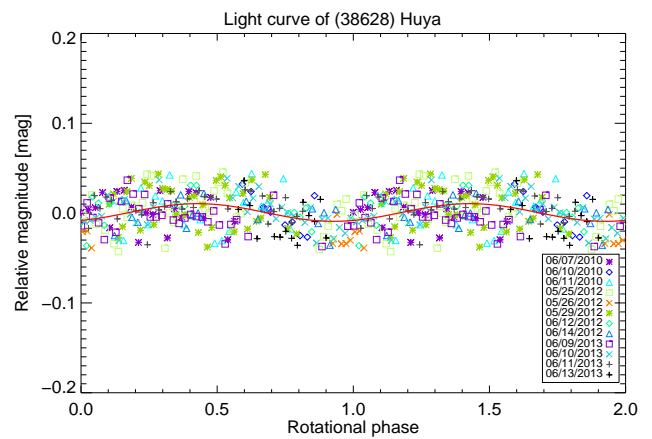}
\caption{\textit{Huya lightcurve}: rotational phase curve for Huya obtained by using a spin period of 5.28~h. The continuous line is a Fourier series fit of the photometric data. Different symbols correspond to different dates.}
\label{fig:LC_Huya}
\end{figure}

Huya was observed in 2010, 2012, and 2013 with the 1.5~m OSN telescope and with the 1.23~m Calar Alto telescope. The Lomb periodogram (Figure~\ref{fig:Lomb_Huya}) shows one peak with a high spectral power located at 5.28~h (4.55~cycles/day), and three aliases located at 6.63~h (3.62~cycles/day), at 4.31~h (5.57~cycles/day), and at 9.15~h (2.62~cycles/day). However, in all cases, these peaks have a lower spectral power than the main peak. All techniques confirm the highest peak at 5.28~h, and the aliases with a lower spectral power. In Figure~\ref{fig:LC_Huya}, the corresponding single-peaked lightcurve with an amplitude of 0.02$\pm$0.01~mag is plotted. We must point out that the rotational period around 6~h noted by \citet{Ortiz2003} is also a possibility in the newest data set, but as an alias. As already mentioned, for very low amplitude objects, it is difficult to estimate a secure rotational period estimation. In fact, small variations in the photometry can transmit more power to/from a 24~h-alias from/to the main peak. Therefore, we cannot discard the alias as a true period.

\section{Discussion}

In this section, we exhaustively study binarity in the Trans-Neptunian belt. The main purpose is to determine the binary and non-binary populations share the same rotational features. 
Finally, we derive several physical properties and propose possible formation models for all binaries whose short-term variability has been reported in this work and in \citet{Thirouin2010, Thirouin2012}. 

\subsection{Derived properties from lightcurves of binary/multiple systems}

The rotational properties of small bodies provide information about important physical properties, such as shape, density, and cohesion \citep{Pravec2000, Holsapple2001, Holsapple2004, Thirouin2010, Thirouin2012}. For binaries it is also possible to derive several physical parameters of the system components, such as diameters of the primary/secondary and albedo under some assumptions. Studying the short-term variability of binary/multiple systems also allows us to identify which systems are tidally locked and which are not \citep{Rabinowitz2013, Benecchi2013}.  

In the next paragraph, we present the methodology for deriving the density, albedo, and primary/secondary sizes from the lightcurve. Then, we compare our results as well as our technique for deriving this information from the lightcurve with other methods. In fact, the density, albedo and/or sizes of both components can also be obtained from other methods, such: i) thermal or thermophysical modeling based on data obtained, for example, with the \textit{Herschel Space Observatory} or the \textit{Spitzer Space Telescope} \citep{Muller2010, Stansberry2008}, ii) from the mutual orbit of the binary component \citep{Grundy2008, Brown2010}, iii) from direct imaging \citep{Brown2004}, or iv) from stellar occultation by (B)TNOs \citep{Sicardy2011, BragaRibas2013}. However, these methods only provide some information which requires the complement of other techniques. For example, thermal modeling, which provides the albedo and effective diameter of the system, requires the absolute magnitude as well as the rotational period of the object to derive reliable results \citep{Muller2010, Lellouch2010, Lim2010, Vilenius2012}. On the other hand, stellar occultations allow us to derive the size of the object with a high precision, but the system density can only be derived if the system mass is known \citep{Sicardy2011}. If the system mass is unknown (or if the object is not a binary), the lower limit of the density can only be estimated from the lightcurve \citep{Ortiz2012Makemake}.  

\subsubsection{Density, size, and albedo from lightcurves: methodology}

Assuming TNOs as triaxial ellipsoids, with axes a$>$b$>$c (rotating along c), the lightcurve amplitude,  \textit{$\Delta{m}$}, varies as a function of the observational (or aspect) angle \textit{$\xi$} (angle between the rotation axis and the line of sight) according to \cite{Binzel1989}:
\begin{equation}
\Delta{m} = 2.5~\log\left( \frac{\it{a}}{\it{b}}\right)  - 1.25~\log\left(
\frac{\it{a}^{2}\cos^{2}\xi + \it{c}^{2}\sin^{2}\xi}{\it{b}^{2}\cos^{2}\xi +
\it{c}^{2}\sin^{2}\xi}\right).
\end{equation}
The lower limit for the object elongation (\textit{a/b}), assuming an equatorial view ($\xi$ = 90$^{\circ}$), is $\Delta${m} = $2.5~\log\left(\frac{a}{b}\right)$.
According to the study of \cite{Chandrasekhar1987} of equilibrium figures for fluid bodies, we can estimate lower limits for densities from rotational periods and the elongation of objects. That is to say, assuming that a given TNO is a triaxial ellipsoid in hydrostatic equilibrium (a Jacobi ellipsoid), we can compute a lower density limit. 
Based on this $\rho$, one can define the volume of the system as \textit{V}$_{system}$ = \textit{M}$_{system}$$/$$\rho$, where \textit{M}$_{system}$ is the mass of the system and is known from the orbit of the system. We assume that both components have the same density, which is the system density. If the two components of the system have the same albedo, the primary radius (\textit{R}$_{primary}$) can be expressed as 
\begin{equation}
$\it{R}$_{primary} = \left( \frac{3 \textit{V}_{system}}{4\pi\left(1+10^{-0.6\Delta_{mag}}\right)}\right)^{1/3},
\label{Eq:Primary}
\end{equation}
where $\Delta_{mag}$ is the component magnitude difference \citep{Noll2008}. Assuming that both components have the same albedo, the satellite radius is $\textit{R}_{satellite} = \textit{R}_{primary}10^{-0.2\Delta_{mag}}$.
 

\begin{tiny}
\begin{table*}
\caption{\label{Summary_photo} Summary of the results from this work and from the literature. In the first part of this table, we report the results obtained in this work. We present the preferred rotational period (rot. per. in hour), the preferred photometric period (phot. per. in hour) and the peak-to-peak lightcurve amplitude ($\Delta$m in magnitude), the Julian Date ($\varphi_{0}$) for which the phase is zero in our lightcurves. The Julian Date is without light-time correction. The preferred photometric period is the periodicity obtained from the data reduction. In one case, as mentioned in the photometric results section, we preferred the double rotational periodicity because of the high-amplitude lightcurve (the preferred rotational period). We also indicate the significance level of the preferred rotational period (SL in percent), and other possible rotational periods that are aliases in our study (aliases, in hours). In the second part on this table, we report the results from the literature (rotational period in hours: rot. per. lit., and lightcurve amplitude in magnitude: $\Delta$m lit.) with their references (Ref).   }
\begin{tiny}
\begin{tabular}{@{}lccclcc||ccc} 
\hline
 
 System  & Phot. per. & Rot. per. & $\Delta$m &  $\varphi_{0}$ [JD]  & SL & Aliases & Rot. per. lit. &  $\Delta$m lit.  &Ref.   \\
 
  & [h] &  [h]& [mag] & [2450000+]  & [$\%$] & [h] &[h]& [mag]& \\
\hline
\hline
 
Salacia &  6.5 & 6.5  & 0.06$\pm$0.02   &  5743.58501 & 99 & 8.86 & 6.09 or 8.1 &  0.03$\pm$0.01 & T10\\
... & ...  &...& ...  &   ... & ... & ...& - & <0.04 & B13\\
Varda & 5.91  & 5.91  & 0.02$\pm$0.01   &     5037.43042   & 95 & 7.87, 4.76 & 5.9 or 7.87 &0.06$\pm$0.02 & T10\\
... & ...  &...& ...  &   ... & ... & ...& - & <0.04 & B13\\
2007~TY$_{430}$  & 4.64  &  9.28  &  0.24$\pm$0.05$^{a}$  &   5863.49277   & 99 & 6.22  & -&  - & -\\
Huya  & 5.28   &5.28  & 0.02$\pm$0.01 & 5355.38744 & 95 & 6.63, 4.31, 9.15 & - & <0.06 & SJ02 \\
...  & ...   &... & ... & ... & ... &...  & - & <0.04 & LL06\\
...  & ...   &... & ... & ... &  ...&  ...& 6.75 or 6.68 or 6.82 & <0.1 & O03\\
2007~UK$_{126}$  &  11.05 &  11.05 & 0.03$\pm$0.01 &  5863.54538     & 95& 14.3, 20.25  & -& -& -\\
2002~VT$_{130}$  &  ? &  ? & 0.21$^{b}$ &  5867.59876    &- & -  &- &  - &-\\

\hline\hline

\end{tabular}
\\
Notes: $^{a}$: Peak-to-peak amplitude of the double-peaked lightcurve; $^{b}$: amplitude variation based on 4~h of observations (see Section~\ref{sec:VT130}); references list: SJ02: \citet{SheppardJewitt2002}; 
O03: \citet{Ortiz2003}; LL06: \citet{LacerdaLuu2006}; T10: \citet{Thirouin2010}; B13:\citet{Benecchi2013}.
 \end{tiny}
\end{table*}
\end{tiny}


We can derive the geometric albedo in the \textit{$\lambda$} band, \textit{p}$_{\lambda}$, given by the equation:
\begin{equation}
$\it{p}$_{\lambda} =  \left(\frac{\textit{C}_{\lambda}}{\textit{R}_{effective}}\right)^{2}10^{-0.4\textit{H}_{\lambda}},
\end{equation}
where \textit{C}$_{\lambda}$ is a constant depending on the wavelength \citep{Harris1998}, and \textit{H}$_{\lambda}$ is the absolute magnitude in the \textit{$\lambda$} band. The effective radius of the system, \textit{R}$_{effective}$, is $\textit{R}_{effective}$ = $\sqrt{\textit{R}_{primary}^{2}+\textit{R}_{satellite}^{2}}$.
It is important to remember that we derived the lower limit of the density, so the derived sizes are upper limits and the derived albedo is a lower limit. We used the same method and the same assumptions to derive the density of Jacobi and MacLaurin objects.

\subsubsection{Density, size, and albedo from lightcurves: results}

In Table~\ref{tab:sizealbedo}, the density, primary and satellite sizes and albedo from lightcurves are summarized for each system studied in this work, in \citet{Thirouin2010}, and in \citet{Thirouin2012}. To estimate the sizes and the albedos using previous equations, we need the system masses that are summarized in Table~\ref{tab:sizealbedo}. From the lightcurves, we report albedo, primary/satellite sizes, and the lower limit of the density for seven systems. In three cases, we are only able to derive the lower limit of the density. For the system 2002~VT$_{130}$, we are not able to derive these parameters because we can only constrain its short-term variability. 

As already pointed out, low-amplitude lightcurves ($\Delta$m$\leq$0.15~mag) can be explained by albedo heterogeneity on the surface of a body that does not have a Jacobi shape, while large-amplitude lightcurves ($\Delta$m$>$0.15~mag) are probably due to the shape of an elongated Jacobi body. In Table~\ref{tab:sizealbedo}, we reported a lower limit of the density computed following the study of \citet{Chandrasekhar1987} of equilibrium figures for fluid bodies. To compute the lower limit of the density according to \citet{Chandrasekhar1987}, we have to assume that the object is a triaxial ellipsoid in hydrostatic equilibrium. This means that, according to our criterion, only objects with a large-amplitude lightcurve have to be considered as Jacobi ellipsoids. In our sample, only two binary systems have a large-amplitude lightcurve: 2007~TY$_{430}$ and 2001~QY$_{297}$, and they can be considered as Jacobi ellipsoids. We found that the 2001~QY$_{297}$ system has a very low lower density limit of 0.29~g cm$^{-3}$, we derived a primary radius of <129~km, a secondary radius of <107~km, and a geometric albedo of >0.08 for the two components. For the system 2007~TY$_{430}$, we found that the two components have similar radii of <58~km (primary) and <55~km (secondary), we derived a lower limit to the density of 0.46~g cm$^{-3}$, and a geometric albedo of >0.12 for both components. Most of the lightcurves of binary systems studied in this work and in \citet{Thirouin2010, Thirouin2012} are more significantly affected by albedo effects than by shape effects. In these cases, the objects are most likely spheroids, not Jacobi ellipsoids. As already pointed out, the lower limit of the density, and other parameters are only very crude estimate.   

\subsubsection{Density, size, and albedo from other methods}

The component sizes and/or albedo can be estimated by other means. It is possible to verify whether the derived parameters from lightcurves and other methods are consistent or not, to check the validity of our method. In Table~\ref{tab:sizealbedo}, the density, the sizes of both components, and/or the albedo derived from other method(s) are summarized. As already mentioned, our method is only valid for Jacobi ellipsoids, threfore care has to be taken for objects with low variability, which are presumably MacLaurin spheroids. We must point out that for non-spherical bodies the concept of radius is unsuitable and we need to talk about an equivalent radius to that of a sphere in volume or in area. In conclusion, one must keep in mind that the radii proposed in \citet{Stansberry2008}, \citet{Lellouch2010}, \citet{Muller2010}, \citet{Mommert2012}, \citet{Vilenius2012}, \citet{Fornasier2013}, and \citet{Vilenius2014} are equivalent radii of the projected area, which means that densities should not be computed from these values. 

For the systems Quaoar-Weywot and Orcus-Vanth, our values are clearly unrealistic, but this was expected because in both cases, we studied MacLaurin spheroids. Based on the low-amplitude lightcurves of the systems Salacia-Actaea, Varda-Ilmar\"{e}, and Typhon-Echidna, we have to expect MacLaurin spheroids. For Salacia-Actaea, we derived a density >1~g cm$^{-3}$, a primary (secondary) radius of <491~km (<165~km), and a geometric albedo of >0.03 for the two components. Our albedo estimate is lower than the albedo obtained with thermal modeling, and we derived higher radii. For the systems Varda-Ilmar\"{e} and Typhon-Echidna, we derived higher albedos and lower radii for the components. However, we must point out that our estimates for these three systems are consistent with the thermal modeling within the error bars. For 2007~TY$_{430}$, there is no study available to confirm our estimations. \citet{Sheppard2012} concluded that assuming a minimum density of 0.5~g cm$^{-3}$, the system albedo is $>$0.17, and that of the radii of the two components are $<$60~km. This is similar to our own results, but they assumed a minimum density to start with. For the system 2001~QY$_{297}$, \citet{Vilenius2014} derived a density of 0.92$^{+1.30}_{-0.27}$~g cm$^{-3}$, a geometric albedo of 0.075$^{+0.037}_{-0.027}$, and primary/secondary radii of about 85~km/77~km. For Huya, 2003~AZ$_{84}$, and 2007~UK$_{126}$, we only derived the lower density limits. However, because all of these systems have a low-amplitude lightcurve, our estimate can be unrealistic. In conclusion, deriving physical parameters such as albedo, sizes, and density from the lightcurves of binary systems is a reliable technique for Jacobi ellipsoids.  

\subsection{Lightcurve amplitude and rotational-period distributions.}

Using the literature and the results presented in this work, we created a database of lightcurves with rotational periods and/or lightcurve amplitudes of binary/multiple systems. This database, updated on December 2013, is presented in Table~\ref{Tab:PeriodAmplitude}. We compiled 32 primaries and 3 satellites with a rotational period and/or peak-to-peak amplitude or constraints \footnote{A recent study of 2010~WG$_{9}$ suggested a rotational period of 131.89$\pm$0.06~h or 263.78$\pm$0.12~h \citep{Rabinowitz2013}. Such long rotational periods have been observed only for tidally evolved BTNOs, suggesting that this object may be such a system. However, as the binarity of this object is not confirmed yet, we did not include it in our study.}. 

\subsubsection{Lightcurve amplitude distributions}

\begin{figure}
 \includegraphics[width=9cm, angle=0]{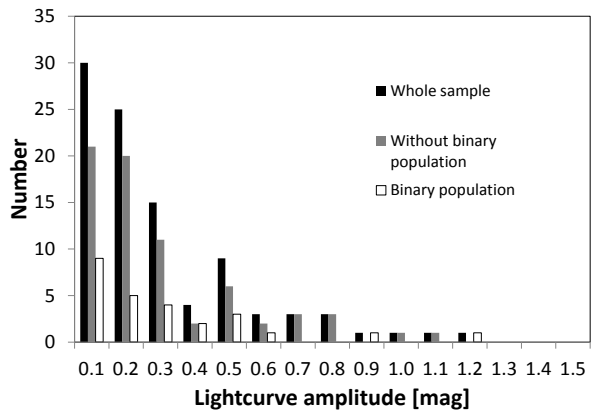}
\caption{\textit{Number of objects versus lightcurve amplitude:} we consider three different samples: the whole sample (black bars), the sample without binary objects (gray bars), and the binary sample (white bars).}
\label{fig:Ampl1}
 \end{figure}

\begin{figure}
\includegraphics[width=9cm, angle=0]{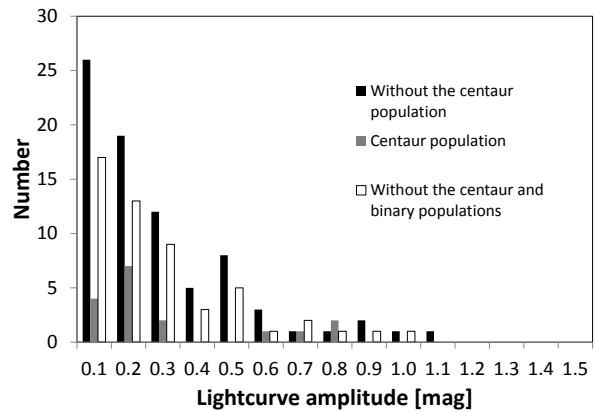}
\caption{\textit{Number of objects versus lightcurve amplitude:} we consider three different samples: the sample without the centaur population (black bars), the centaur population (gray bars), and the sample without the centaur and the binary populations (white bars).}
\label{fig:Ampl2}
 \end{figure}

The number of objects with a lightcurve amplitude value reported in the literature and in this work are reported in Figs.~\ref{fig:Ampl1} and \ref{fig:Ampl2}. Objects with only a constraint of their lightcurve amplitude were not taken into account. 

In the Figure~\ref{fig:Ampl1} we focus on three samples: the entire sample, the binary population, and the sample without the binary population. First of all, we must point out that most objects (in the three samples) have an amplitude of $<$0.2~mag. About 57$\%$ of the entire sample, 59$\%$ of the sample without the binary population, and 54$\%$ of the binary sample have a low amplitude.   
The main reason for observing nearly flat lightcurves is probably a spherical object (or MacLaurin) with low albedo variations along the surface. The second option for this lightcurve amplitude can be an overabundance of nearly pole-on orientation of the objects (rotational axis toward the observer). The most reasonable option is to consider that the observed objects are mostly MacLaurin spheroids with albedo marks with low contrast on their surfaces. \citet{Duffard2009} presented a model that used a Maxwellian rotational frequency distribution such as that obtained previously in this work, and assumed that the objects adopt hydrostatic equilibrium shapes. The model generates a set of 100,000 objects, and each object is randomly assigned to a rotation period from the distribution. All objects are assumed to be in hydrostatic equilibrium with a fixed density. The lightcurve amplitude is only a result of the shape of the body
and the inclination of its rotation axis (randomly chosen). The body shapes are computed using \citet{Chandrasekhar1987} equations. In other words, Jacobi ellipsoids produce a non-flat lightcurve, whereas MacLaurin spheroids generate a nearly flat lightcurve. One of the results of this model is that for a fixed density of 1~g cm$^{-3}$, one expects 55.63$\%$ of MacLaurin spheroids and only 12.61$\%$ of Jacobi ellipsoids, while for a fixed density of 1.5~g cm$^{-3}$, one expects 11.92$\%$ of Jacobi ellipsoids and 72.31$\%$ of MacLaurin spheroids. Because the centaurs are small and are not expected to be in hydrostatic equilibrium, and moreover, their rotations may be more evolved than those of the pure TNOs, we have removed them from the sample in Figure~\ref{fig:Ampl2}. 

The sample without the binary and the centaur populations mainly has amplitudes between of 0.1 and 0.2~mag. In conclusion, there are hints that the binary amplitudes may be slightly larger than the non-binary population, but overall the distributions are similar, and only more studies about the short-term variability of binary systems will allow us to confirm or refute this tendency. 

\subsubsection{Rotational period distributions}

\citet{Salo1987} showed that asteroid rotation rates probably follow a Maxwellian distribution if the asteroids are in collisional equilibrium.
\citet{Binzel1989} studied the asteroid rotation rate distributions and concluded that for asteroids with a diameter \textit{D}>125~km, a Maxwellian is able to fit the observed rotation rate distributions, implying that their rotation rates may be determined by collisional evolution. In contrast, for asteroids with a diameter \textit{D}<125~km, there is an excess of slow rotators and their non-Maxwellian distributions suggests that their rotation rates are more strongly influenced by other processes that result from their formation in catastrophic disruption events, et cetera. However, a recent study by \citet{Warner2009} suggested that the non-Maxwellian distribution for small main belt asteroids (MBAs) is mainly due to the Yarkovsky–O'Keefe–Radzievskii–Paddack (YORP, see \citet{Rubincam2000} for more details) effect. Although TNOs are very different from asteroids, we can carried out a similar study. Because the number of TNOs/centaurs whose short-term variability has been studied is still too limited, we did not divide the sample according to object sizes, and moreover, all TNOs studied here are larger than 125~km. As pointed out in \citet{Binzel1989}, there are several biases in the asteroid lightcurve database, mainly because it is easier to determine reliable and publishable parameters for asteroids that have short rotational periods with a large lightcurve amplitude. Similar biases have been noted in the TNOs/centaurs lightcurve database \citep{Sheppard2008, Thirouin2010}. \citet{Binzel1989} tried to effectively eliminate bias effects by including all asteroids, even those with a low reliability code (a low reliability code means that the estimated rotational period has a low confidence level and may be incorrect). \citet{Binzel1989} stressed that excluding low reliability objects results in overweighting asteroids with large amplitudes and short periods, which introduces a significant bias in the results of the statistical studies. Based on such a study, we decided to proceed in the same way and included all TNOs/centaurs with a short-term variability study, even if the rotational period estimated was not unambiguously determined. Finally, the bin size used here for the histograms is the same as that used in \citet{Binzel1989}, mainly because the sample of TNOs/centaurs with a short-term variability study is still too small to consider smaller bin sizes. 

For multiple determinations of the period and/or amplitude, we selected the value preferred by the author(s) who published the study. If no preferred value was mentioned, we proceeded to a random choice. 
In some cases several rotational periods have been proposed for an object, for these we randomly chose one of these rotational periods. Either this, or one can simply give weights to each period according to the number of possible periods. 

As in \citet{Binzel1989}, we fitted the rotational frequency distribution to a Maxwellian distribution expressed as   
\begin{equation}
\textit{f}~({\Omega}) = \sqrt{\frac{2}{\pi}} {\frac{\textit{N} \Omega^{2}} {\sigma^{3}}} exp \left( \frac{-\Omega^{2}} {2\sigma^{2}}\right), 
\end{equation}
where N is the number of objects, \textit{$\Omega$} is the rotation rate in cycles/day, \textit{$\sigma$} is the width of the Maxwellian distribution. The mean value of this distribution is $\Omega_{mean} = \sqrt{\frac{8}{\pi}} \sigma $.

\begin{figure} 
\includegraphics[width=9cm, angle=0]{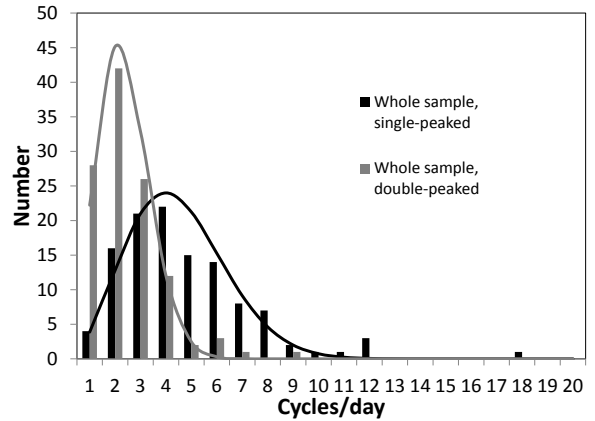}
\caption{\textit{Number of objects versus rotational rate in cycles/day:} two different samples are plotted. Black bars: the entire sample considering that all the lightcurves are single-peaked (number of objects (N)=115). Gray bars: the entire sample considering double-period lightcurves. A Maxwellian fit to the entire sample assuming a single period distribution gives a mean rotational period of 5.35~h ($\sigma$=2.813$\pm$0.262, continuous black line). The Maxwellian fit of the second sample gives a mean rotational period of 10.11~h ($\sigma$=1.488$\pm$0.139, continuous gray line). } 
\label{fig:Hist0}
\end{figure}

\begin{figure}
\includegraphics[width=9cm, angle=0]{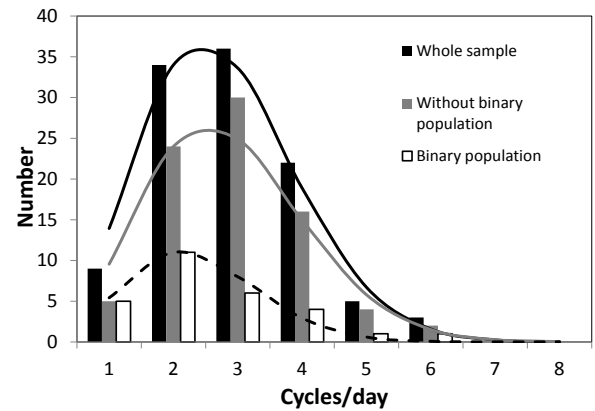}
 \caption{\textit{Number of objects versus rotational rate in cycles/day:} three different samples are plotted: the entire sample (number of objects (N)=109, black bars), the binary population (N=28, white bars), and the sample without the binary population (N=81, gray bars). A Maxwellian fit to the entire sample gives a mean rotational period of 8.63~h ($\sigma$=1.743$\pm$0.167, continuous black line). The Maxwellian fit of the sample without binary objects gives a mean rotational period of 8.37~h ($\sigma$=1.796$\pm$0.199, continuous gray line). Finally, the fit for the binary population gives a mean rotational period of 10.11~h ($\sigma$=1.487$\pm$0.281, dashed black line). }
\label{fig:Hist1}
\end{figure}

For many TNO lightcurves it is extremely difficult to distinguish whether the variability is due to albedo variations or an elongated shape. In other words, the derived rotational period have an uncertainty of a factor 2. In fact, a double-peaked lightcurve is expected for a non-spherical object since the projected cross-section should have two minima and two maxima along one full object rotation. In contrast, lightcurves caused by albedo variations on the surface of the object are expected to be single-peaked (see \citet{Lacerda2003} for a complete review). Figure~\ref{fig:Hist0} was plotted considering that all objects (TNOs and centaurs) have i) single-peaked lightcurves and ii) double-peaked lightcurves. These are obviously extreme cases that are unrealistic. From Maxwellian fits to the rotational frequency distributions, the mean rotational periods are 5.35$\pm$0.31~h for the entire sample assuming a single-peaked rotation distribution, and 10.11$\pm$0.59~h for the second distribution. Obviously, both distributions are unrealistic, but an upper limit to the rotational period distribution is provided assuming the double distribution. A more realistic distribution is to assume that all the lightcurves with an amplitude lower than 0.15~mag are single-peaked (i.e., assuming that variability is due to albedo features), and lightcurves with a variability higher than 0.15~mag are double-peaked (i.e., assuming that variability is due to the object's elongated shape). In Figure~\ref{fig:Hist1} we show such a distribution. Obviously, this criterion is a good first approximation, but some cases do not fit in it. For example, a double-peaked rotational period of 5.92~h and a lightcurve amplitude of 0.09~mag have been reported for the Centaur Chiron \citep{Bus1989}. According to our criterion, a single-peaked lightcurve has to be considered, but a rotational period of 2.96~h is quite unrealistic for this kind of object. In conclusion, in some cases it is necessary to decide on a case-by-case basis.  

In Figure~\ref{fig:Hist1}, three different samples are plotted: the entire sample, the binary population, and the sample without the binary population. From Maxwellian fits to the rotational frequency distributions, the mean rotational periods are 8.63$\pm$0.52~h for the entire sample, 8.37$\pm$0.58~h for the sample without the binary population and, finally, 10.11$\pm$1.19~h for the binary population. 
 
\begin{figure}
\includegraphics[width=9cm, angle=0]{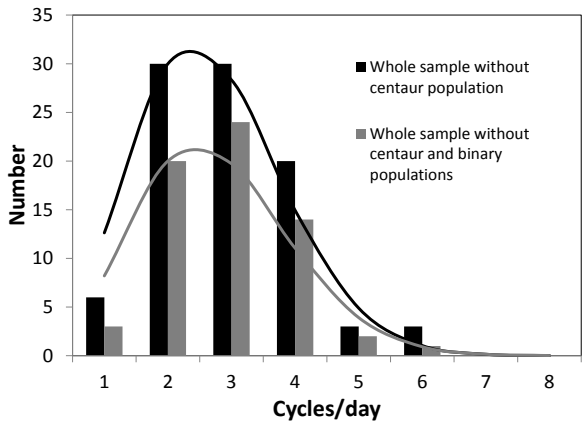}
 \caption{\textit{Number of objects versus rotational rate in cycles/day:} two different samples are plotted: the sample without the centaur population (number of objects (N)=92, black bars), and the sample without the centaur and binary populations (N=64, gray bars). A Maxwellian fit to the first sample gives a mean rotational period of 8.86~h ($\sigma$=1.697$\pm$0.177, continuous black line). The second Maxwellian fit of the sample without binaries nor centaurs gives a mean rotational period of 8.64~h ($\sigma$=1.741$\pm$0.217, continuous gray line).  
 }
\label{fig:Hist2}
\end{figure}

\citet{Duffard2009} noted that the centaur population has a higher mean rotational period. In fact, as this population is more collisionally evolved, their rotational periods might be affected. We removed them from our different samples and obtained two new Maxwellian distributions. In Figure~\ref{fig:Hist2}, we plot the sample without the centaur population and the sample without the centaur and the binary populations. Based on the Maxwellian distribution fits, we computed a mean rotational period of 8.86$\pm0.58$~h, 8.64$\pm$0.67~h for the sample without the centaur population and for the sample without centaur and the binary populations, respectively. 

In conclusion, based on the Maxwellian distribution fits, we found a mean rotational period for the sample without the binary and the centaur populations of 8.64$\pm$0.67~h, whereas the binary population seems to have a higher mean rotational period of 10.11$\pm$1.19~h. This means that the binary systems rotate more slowly. We must point out that the number of binary/multiple systems whose short-term variability has been studied is limited, but it is reasonable to expect that binary systems have longer rotational periodicities. In fact, tidal effects are able to slow down the primary (and the secondary) rotational rate (see Section 5.3). 

\subsection{Tidal effects}

Tidal effects can synchronize the spin rate of the primary and/or of the secondary to its orbital period, and can circularize the satellite orbit. 

\subsubsection{Circularization time}

According to \citet{Goldreich2009}, the tides raised on the primary are
\begin{equation}
\frac{1}{\textit{e}} \frac {d\textit{e}}{d\textit{t}}= \frac{57}{8} \frac{\textit{k}_{primary}}{\textit{Q}_{primary}} \frac{\textit{M}_{satellite}}{\textit{M}_{primary}} \left(\frac{\textit{R}_{primary}}{\textit{a}}\right)^5 \textit{n}, 
\label{Eq:Circular} 
\end{equation}
where \textit{e} is the orbital eccentricity, \textit{M}$_{satellite}$ and \textit{M}$_{primary}$ are the satellite and primary masses (respectively), \textit{a} is the orbital semimajor axis, \textit{Q}$_{primary}$ is the dissipation parameter of the primary, \textit{R}$_{primary}$ is the primary radius, and \textit{n} is the mean orbital angular velocity. The dissipation parameter depends on the body rigidity, the acceleration of gravity at the object surface, density, and size. According to \citet{Goldreich1966}, this parameter range is 10 to 6$\times$10$^{4}$. The typical value used in the TNO case is \textit{Q}=100 \citep{Noll2008}.  
The Love parameter of the primary is
\begin{equation}
\textit{k}_{primary} =  {1.5 \over 1 + 19 \mu _{primary} /( 2 \rho _{primary} \textit{g}_{primary} \textit{R}_{primary})},
\label{Eq:LoveNumber}
\end{equation}
where g$_{primary}$ = GM$_{primary}$/R$_{primary}^{2}$ is the surface gravity of the primary, $\rho_{primary}$ is the primary density, and $\mu_{primary}$ is the primary rigidity. The rigidity is estimated to be 3$\times$10$^{10}$ N~m$^{-2}$ for rocky objects and 4$\times$10$^{9}$ N~m$^{-2}$ for icy ones \citep{Gladman1996}.  
In Table~\ref{Tab:tidallockgladman}, all the BTNOs whose short-term variability was studied in this work, and in \citet{Thirouin2010, Thirouin2012} are reported, as well as the parameters needed to compute the circularization time. For Actaea, Ilmar\"{e}, and the satellite of Huya, the times required to circularize the orbit are short or similar to the age of the solar system, and therefore we expect nearly circular orbits. With an orbital eccentricity of 0.0084$\pm$0.0076 and 0.02$\pm$0.04 for Actaea and for Ilmar\"{e} (respectively), both orbits are nearly circular \citep{Stansberry2012, Grundy2011}. The orbit of Huya's satellite is unknown, but we have to expect a nearly circular orbit. The times required to circularize the orbit of Echidna and the satellite of 2007~UK$_{126}$ are long, therefore we have to expect non-circular orbits. With an orbital eccentricity of 0.526$\pm$0.015 for Echidna, its orbit is not circular \citep{Grundy2008}. The orbit of the satellite of 2007~UK$_{126}$ is unknown, but we expect a non-circular orbit. The orbit of the satellite of 2007~TY$_{430}$ is far from circular and will require a long time to be circular. According to \citet{Sheppard2012}, the orbital eccentricity is 0.1529$\pm$0.0028, which confirms the non-circular orbit. The orbits of Weywot and of the satellites of 2001~QY$_{297}$ and 2003~AZ$_{84}$ will also require a long time to be circular. \citet{Fraser2013} derived an orbital eccentricity of $\sim$0.13-0.16 for Weywot, and \citet{Grundy2011} estimated an orbital eccentricity of 0.4175$\pm$0.0023 for the satellite of 2001~QY$_{297}$. This means that neither orbit is circular. The orbit of the satellite of 2003~AZ$_{84}$ is unknown, but we expect a non-circular orbit. Based on the time required to circularize the orbit of Vanth, we can expect a non-circular orbit, which disagree with the upper limit of the eccentricity of 0.0036 estimated by \citet{Brown2010}. But as pointed out in \citet{Ortiz2011}, it is quite possible that Vanth has a much larger mass and size than originally estimated by \citet{Brown2010}. However, not only the tidal effect may circularize the orbit, the Kozai mechanism can do this as well \citep{Kozai1962}. Owing to the perturbation by the Sun, the orbit of the satellite experiences libration (oscillation) of its argument of pericenter. \citet{Porter2012} presented an exhaustive study about the Kozai effect on BTNOs. They simulated a large set of synthetic BTNOs and confirmed that the Kozai effect can completely reshape the initial orbits of the systems. According to the simulations of \citet{Porter2012}, we have to expect many BTNOs with a very tight and circular orbits. To date, only 21 objects have well-known orbits and 30 objects have ambiguous orbits, but apparently several systems have near-circular orbits \citep{Grundy2011DPS}.  

 
\subsubsection{Synchronization time}

Tidal effects can also synchronize the satellite and primary spin rates to the orbital period. Several formulas have been proposed in the literature to estimate the time needed to lock the primary/secondary rotational rates \citep{Hubbard1984, Gladman1996}. Here, we compute such a time using the formula of \citet{Gladman1996} which takes into account the body rigidity.  
According to \citet{Gladman1996}, the time needed ($\tau_{lock}$) to tidally lock a primary is expressed as
\begin{equation}
\tau _{lock} = \frac {\omega _{primary} \textit{a}^6 \textit{I}_{primary} \textit{Q}} {3 \textit{G} \textit{M}^2_{satellite} \textit{k}_{primary} \textit{R}_{primary}^5}, 
\label{Eq:LockGladman}
\end{equation}
where $\omega _{primary}$ is the initial rotational rate of the primary, \textit{a} is the distance between the primary and the satellite, \textit{Q} is the dissipation, \textit{G} is the gravitational constant, \textit{M}$_{satellite}$ and \textit{R}$_{primary}$ are, respectively, the mass of the satellite and the radius of the primary, \textit{I} is the moment of inertia of the primary (such as I=0.4M$_{primary}$R$_{primary}^{2}$), and \textit{k}$_{primary}$ is the Love number of the primary. 

In Table~\ref{Tab:tidallockgladman} we summarize the parameters used to compute the time needed to tidally lock the primary. We considered three cases: i) the density of the satellite is the same as the density of the primary (i.e., the system density), ii) the density of the satellite is 1 g~cm$^{-3}$, and iii) the density of the satellite is 0.5 g~cm$^{-3}$. 
Assuming a dissipation of Q=100 and that both components have the same density, times to tidally lock most of the binaries presented here are long (compared with the age of the solar system), and so we have to expect that none of these systems is tidally locked. We must point out that this fact is confirmed thanks to our short-term variability studies of these systems, which show evidence for rotation periods of several hours (see \citet{Thirouin2010, Thirouin2012}, and this work). The times to tidally lock Huya and Varda are short. Therefore, we can expect that these systems can be tidally locked. However, there is evidence for rotation periods of several hours, and therefore primaries are not tidally locked (see \citet{Thirouin2010, Thirouin2012}, and this work). By considering a satellite with a lower density (0.5 g~cm$^{-3}$) and a rigidity for rocky bodies, we computed times to tidally lock the primary at about 10$^{9}$~years; a higher Q might also explain longer locking times. The computed tidal locking times according to \citet{Gladman1996} seem agree with our observational results. In conclusion, the systems studied here and in \citet{Thirouin2010, Thirouin2012} are not yet synchronous (or double synchronous). But the tidal effects between the primary and the satellite might already have slowed down the primary rotational rate and might explain the rotational period distributions. On the other hand, we must point out that tidal circularization and tidal despinning are complex effects. For example, for equal-size objects, the secondary tidal effect cannot be neglected. The assumptions used to derive Equation~\ref{Eq:LockGladman} are not valid for binaries with a moderate to high eccentricity. More studies about tidal effects as well as estimations of the parameter Q are needed. 

\subsection{Formation of binary and multiple systems}

Various models have been proposed to explain the formation of binary and multiple systems, such as the L$^{3}$ mechanism by \citet{Goldreich2002}, the L$^{2}$s mechanism by \citet{Goldreich2002}, the chaos-assisted capture by \citet{Astakhov2005}, the collisional model by \citet{Durda2004}, the hybrid model by \citet{Weidenschilling2002}, the gravitational collapse by \citet{Nesvorny2010}, and the rotational fission model by \citet{Ortiz2012}. A complete review of some of these mechanisms can be found in \citet{Noll2008}.

\begin{figure}
 \includegraphics[width=9.5cm, angle=0]{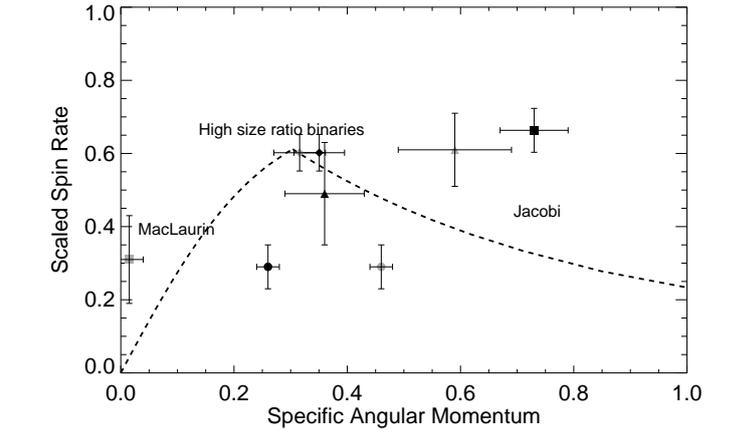}
\caption{\textit{Scaled Spin Rate versus specific Angular Momentum}: scaled spin rate and specific angular momenta computed as mentioned in the text. We indicate the MacLaurin and the Jacobi sequences. The "high size ratio binaries", as indicated in \citet{Descamps2008},are located near the MacLaurin/Jacobi transition. The legend is as follows: the black triangle denotes Salacia-Actaea, the gray diamond Haumea-Namaka, the black diamond Haumea-Hi'iaka, the gray triangle Varda-Ilmar\"{e}, the gray square Quaoar-Weywot, the black circle Orcus-Vanth assuming a secondary-to-primary mass ratio of 0.03, the gray circle Orcus-Vanth assuming a secondary-to-primary mass ratio of 0.09, and black square for Typhon-Echidna. The system 2007~TY$_{430}$ is not plotted here because we restricted the plot for a better visualization). Error bars are approximate. }
\label{fig:SAM}
 \end{figure}

Both capture and collisional models require that the number of TNOs in the primordial Trans-Neptunian belt was at least several orders of magnitude higher than currently and consequently, BTNOs are primordial systems \citep{Petit2004}. Only the formation of a few binary systems is well known, such as the Pluto/Charon formation. In fact, it is complicated to favor or discard any model, especially if the orbit is unknown. Currently, the binary formation via capture and/or collision as well as gravitational collapse are the most often  investigated and seem the most probable in the Trans-Neptunian belt. In fact, the rotational fission scenario is unlikely for most of the binaries, but in some cases it has to be considered \citep{Ortiz2011, Ortiz2012}. One argument in favor of a rotational fission scenario for some cases is the specific angular momentum of a binary and multiple system. The specific angular momentum (SAM), computed according to \citet{Descamps2008} is 

\begin{multline}
\textit{SAM} =\frac{\textit{q}}{(1+\textit{q})^{\frac{13}{6}}}\sqrt{\frac{\textit{a}(1-\textit{e}^{2})}{\textit{R}_{primary}}} + \frac{2}{5} \frac{\lambda_{primary}}{(1+\textit{q})^{\frac{5}{3}}} \Omega + \\
\ \frac{2}{5} \lambda_{satellite} \frac{\textit{q}^{\frac{5}{3}}}{(1+\textit{q})^{\frac{7}{6}}}\left (\frac{\textit{R}_{primary}}{\textit{a}} \right)^{\frac{3}{2}}, 
\end{multline}
where \textit{q} is the secondary-to-primary mass ratio, \textit{a} is the semimajor axis, \textit{e} is the eccentricity, and \textit{R}$_{primary}$ is the primary radius.
The $\Omega$ parameter is the normalized spin rate expressed as $\Omega = \frac{\omega_{primary}}{\omega_{critical}}$, where $\omega_{primary}$ is the primary rotation rate and $\omega_{critical}$ the critical spin rate for a spherical body, such as $\omega_{critical} = \sqrt{\frac{GM_{system}}{{R_{effective}^3}}}$, where \textit{G} is the gravitational constant, \textit{M}$_{system}$ is the system mass and \textit{R}$_{effective}$ the effective radius of the system (or equivalent radius).

Assuming triaxial objects with semi-axes as $ a> b > c$, the $\lambda$ shape parameter is $\lambda_{primary}= \frac{1+\beta^{2}}{2(\alpha\beta)^\frac{2}{3}}$, where $\alpha$ = $c/a$ and $\beta$ = $b/a$. We considered the satellites to be spherical bodies, so $\lambda_{satellite}$=1. 
The scaled spin rate (SSR), according to \citet{Chandrasekhar1987}, is expressed as, $\textit{SSR}= \frac{\omega_{primary}}{\sqrt{\pi G \rho_{primary}}}$, where $\rho_{primary}$ is the density of the primary. We considered that both components have the same density, which is the system density. The scaled spin rate and specific angular momentum are dimensionless values. 

Based on a binary asteroid population study, \citet{Descamps2008} concluded that binary systems near the MacLaurin/Jacobi transition are most likely formed by rotational fission or mass shedding. In Figure~\ref{fig:SAM} we indicate the MacLaurin and Jacobi sequences, the binary/multiple systems studied in this work, and those from \citet{Thirouin2010, Thirouin2012}.  

We computed a specific angular momentum of 0.36$\pm$0.07 and a scaled spin rate of 0.49$\pm$0.14 for the Salacia-Actaea system. These values allow us to discard a rotational fission scenario to explain the formation of this system (Figure~\ref{fig:SAM}). However, we must point out that several of the parameters used to compute the specific angular momentum and the scaled spin rate present a high uncertainty, and considering the error bars, Salacia-Actaea may have suffered a rotational fission. For example, the Salacia-Actaea system density presents a high uncertainty: i) 1.16$^{+0.59}_{-0.36}$~g cm$^{-3}$ according to \citet{Stansberry2012}, ii) 1.38$\pm$0.27~g cm$^{-3}$ according to \citet{Vilenius2012}, and iii) 1.29$^{+0.29}_{-0.23}$~g cm$^{-3}$ according to \citet{Fornasier2013}. The Salacia-Actaea lightcurve is flat, and accordingly, this object presents a homogeneous shape little or no deformation. For this reason and because of the size of the satellite, a collisional scenario is not favored to explain the Actaea formation (except for a Pluto/Charon like formation). We suggest a capture or gravitational collapse model. A possible rotational fission scenario has to be confirmed. 

We computed a specific angular momentum of 0.59 and a scaled spin rate of 0.61 for the Varda-Ilmar\"{e} system (Figure~\ref{fig:SAM}). In this case, we did not compute the error bars \footnote{In Figure~\ref{fig:SAM}, we use an error bar of $\pm$0.1 for the specific angular momentum and the scaled spin rate.} of the specific angular momentum and the scaled spin rate, mainly because no density estimation is available for this system instead we used the lower limit of the density derived in this work, which, as already mentioned, is only a very crude estimation. The lightcurve of the Varda-Ilmar\"{e} system is flat. This means that this object is probably a MacLaurin spheroid (or near) with a limited shape deformation. Because of this, we favor a capture scenario or gravitational collapse instead of a collisional scenario to explain the satellite formation. The second argument to discard a collisional scenario is the size of the satellite. In fact, the large size of the satellite suggests a non-collisional formation, unless it was created in a similar Pluto/Charon formation model. We must mention that a flat lightcurve can be caused by a pole-on orientation. In this case, the object may be deformed, but we cannot detect it in the lightcurve variation. However, we used the lower limit to the density to derive the specific angular momentum and the scaled spin rate, therefore we must keep in mind a possible rotational fission scenario to explain the formation of this system.

The wide binary 2007~TY$_{430}$ with a specific angular momentum around 4.33 and a scaled spin rate around 0.61 is not plotted in Figure~\ref{fig:SAM} because it is out of the scale. To compute the specific angular momentum and the scaled spin rate of this system, we had to use the lower limit of the density derived in this work, which, as already mentioned, is only a very crude estimation. \citet{Sheppard2012} previously discussed in detail all possible (or impossible) formation models for this system. They considered two plausible scenarii: the L$^{3}$ mechanism based on gravitational capture proposed by \citet{Goldreich2002}, and the gravitational collapse mechanism studied by \citet{Nesvorny2010}. 

We computed a specific angular momentum of 0.15$\pm$0.02 and a scaled spin rate of about 0.31$\pm$0.12 for the Quaoar-Weywot. This system does not seem to come from a rotational fission scenario (Figure~\ref{fig:SAM}). The lightcurve of Quaoar-Weywot has a moderate lightcurve amplitude. This means that this object is probably a MacLaurin spheroid (or near) with a limited shape deformation. However, the satellite, Weywot, has a small diameter of 81$\pm$11~km according to \citet{Fornasier2013}, and therefore a collisional scenario seems the best option to explain the satellite formation.

The specific angular momentum of 2001~QY$_{297}$ is 1.85$\pm$0.39 and its scaled spin rate is 0.58$\pm$0.21, which is out of the scale in Figure~\ref{fig:SAM}. These values seem to indicate that the 2001~QY$_{297}$ binary system was not formed by rotational fission. In fact, the high value of the specific angular momentum and the scaled spin rate of this system do not fall into the ``high size ratio binaries" region indicated in Figure~\ref{fig:SAM} of \citet{Descamps2008} and is far from the Jacobi or MacLaurin sequences. Accordingly, we can probably discard a possible rotational fission origin for this binary. We cannot favor any other formation scenario; this asynchronous binary could have been formed by capture and/or collision, or by gravitational collapse.    

We computed a specific angular momentum of 0.73$\pm$0.06 and a scaled spin rate of 0.66$\pm$0.16 for Typhon-Echidna. This is not too far from the high mass ratio binaries that most likely come from fissions. Therefore, we cannot discard rotational fission to explain the system (Figure~\ref{fig:SAM}). The rotational period of the Typhon-Echidna system is not secure, but we can affirm that the lightcurve amplitude is low. This means that this object is probably a MacLaurin spheroid with a limited shape deformation. We favor a capture scenario or gravitational collapse over a collisional scenario to explain the satellite formation. We must mention that flat lightcurve can be caused by a pole-on orientation. In this case, the object may be deformed, but we cannot detect it in the lightcurve variation.  

For Orcus-Vanth, we computed a specific angular momentum of 0.26$\pm$0.02 and a scaled spin rate of 0.29$\pm$0.06 considering a secondary-to-primary mass ratio of 0.03 \citep{Brown2010}, a specific angular momentum of 0.46$\pm$0.02, and a scaled spin rate of 0.29$\pm$0.06 \citep{Ortiz2011}, considering a secondary-to-primary mass ratio of 0.09 (Figure~\ref{fig:SAM}). \citet{Ortiz2011} showed that the satellite rotation is synchronous (the rotational period of the satellite and the orbital period are the same), and that the system is not double-synchronous because the primary is spining much faster than the orbital period \citep{Thirouin2010}. If we assume that the initial spin period of Orcus was approximately its critical value, the total angular momentum lost by the slowdown to 10~h would have been gained by the satellite, which would have reached exactly its current configuration if the mass ratio of the system was about 0.09 (the value obtained by assuming that the Vanth albedo is smaller than that of Orcus, which is probably the case judging from their very different spectra \citep{Carry2011}). This would support the idea that the satellite might be the result of a rotational fission \citep{Ortiz2011}. 

For Huya, we favor a capture scenario or a gravitational collapse because of the satellite size and the flat lightcurve. We propose a very flat lightcurve for 2007~UK$_{126}$, which seems to discard the collisional scenario. However, the size of the satellite is compatible with a collisional formation. In conclusion, we cannot favor or discard any formation model based on our study for this object. Based on only a few hours of observations, 2002~VT$_{130}$ seems to have a high amplitude lightcurve which means that a collisional scenario may be an option. The system 2003~AZ$_{84}$ is composed of a large primary and a small satellite. This means that a collisional scenario seems the best option. 

\section{Summary and conclusions}

We have analyzed the short-term variability of several Binary Trans-Neptunian Objects (BTNOs). Only one object in our sample, 2007~TY$_{430}$, has a high-amplitude lightcurve ($\Delta {m}$$>$0.15~mag) and can be considered to be a Jacobi ellipsoid. Assuming that this system is in hydrostatic equilibrium, we derived a lower limit to the density ($\rho$>0.46 g~cm$^{-3}$), a primary (secondary) radius of <58 (<55~km), and a geometric albedo of 0.12 for the two components. Other BTNOs studied in this work showed small peak-to-peak amplitude variations and are oblate (MacLaurin spheroid). For them we were only able to derive mere academic guesses on density and geometric albedo. But we showed that deriving several parameters from the lightcurves is a reliable method for Jacobi ellipsoids. 

An exhaustive study about short-term variability as well as derived properties from lightcurves allowed us to draw some conclusions for the Trans-Neptunian belt binary population. Based on Maxwellian fit distributions of the spin rate, we suggested that the binary population rotates slower than the non-binary one. This slowing-down can be attributed to tidal effects between the satellite and the primary, as expected. We showed that no system in this work is tidally locked, but the primary de-spinning process may have already affected the primary rate (as well as the satellite rotational rate). We computed the time required to circularize and tidally lock the systems. We used the \citet{Gladman1996} formula to compute the time required to tidally lock the systems, but this formula is based on several assumptions and approximations that do not always hold. The computed times are reasonable in most cases and confirm that none of the systems is tidally locked, assuming that the satellite densities are low ($\lesssim$0.5~g cm$^{-3}$) and have a high rigidity or have a Q>100 (higher dissipation than usually assumed). However, more studies are necessary to understand the tidal effect between primary and satellite, especially for equal-size systems. 


Finally, by studying the specific angular momentum of the sample, we proposed possible formation models for several BTNOs with short-term variability. In several cases, we obtained hints of the formation mechanism from the angular momentum, but for other cases we do not have enough information about the systems to favor or discard a formation model.

  %
 
\section*{Acknowledgments}

We thank the referee for his or her careful reading of this paper. We are grateful to the Sierra Nevada Observatory, Telescopio Nazionale Galileo and CAHA staffs. This research was based on data obtained at the Observatorio de Sierra Nevada, which is operated by the Instituto de Astrof\'{i}sica de Andaluc\'{i}a, CSIC. Other results were obtained at the Telescopio Nazionale Galileo. The Telescopio Nazionale Galileo (TNG) is operated by the Fundaci\'{o}n Galileo Galilei of the Italian Istituto Nazionale di Astrofisica (INAF) on the island of La Palma in the Spanish Observatorio del Roque de los Muchachos of the Instituto de Astrof\'{i}sica de Canarias. This research is also based on observations collected at the Centro Astron\'{o}mico Hispano Alem\'{a}n (CAHA) at Calar Alto, operated jointly by the Max-Planck Institut fur Astronomie and the Instituto de Astrof\'{i}sica de Andaluc\'{i}a (CSIC). A. Thirouin, N. Morales, and J.L. Ortiz were supported by AYA2008-06202-C03-01, and AYA2011-30106-C02-01, which are two Spanish MICINN/MEC projects. A. Thirouin, N. Morales, and J.L. Ortiz also acknowledge the Proyecto de Excelencia de la Junta de Andaluc\'{i}a, J.A.2007-FQM2998. FEDER funds are also acknowledged.

%


\bibliographystyle{aa}
\bibliography{biblio}





\clearpage
\begin{scriptsize}
\begin{onecolumn}
\begin{longtable}{lccccc}
\caption{\label{Tab:PeriodAmplitude} 
Short-term variability of BTNOs. For multiple rotational periods, the preferred rotational period, according to the authors of each study, is indicated in bold.}\\
\hline
\\

Object &   Single-peak periodicity [h] & Double-peak periodicity [h]& Amplitude [mag] & Abs mag. & Ref \\ 
\hline
\hline
{(134340) Pluto }  &  {{ 153.2 }} & { -} & { 0.33} & { -0.7} & { B97}\\
{Charon }&    { 153.6 } & { -} & { 0.08} & { 0.9} & { B97}\\
(148780) Altjira  &   { -} & { -} & { $<$0.3 } & { 5.6} & { S07}\\
(66652) Borasisi   &  { -} & { -} & { $<$0.05} & { 5.9} & { LL06}\\
 &  { 6.4$\pm$1.0} & { -} & { 0.08$\pm$0.02} & {...} & { K06b}\\
(65489) Ceto   &   { -} & { 4.43$\pm$0.03} & { 0.13$\pm$0.02} & { 6.3} & {D08}\\
(136199) Eris  &   { 13.69/28.08/32.13} & { -} & { $<$0.1$\pm$0.01} & { -1.2} & { Du08}\\
&    { 3.55} & { -} & { $\sim$0.5} & { ...} & { L07}\\
 &   { -} & { -} & { $<$0.01} & { ...} & { R07, S07}\\
 &   { 25.92} & { -} & { 0.1} & { ...} & { R08}\\
(136108) Haumea  &  { -} & { 3.9154$\pm$0.0002} & { 0.28$\pm$0.02} & { 0.2} & { R06}\\
 &    { -} & { 3.9155$\pm$0.0001} & { 0.29$\pm$0.02} & { ...} & { L08}\\
 &    { -} & { 3.92} & { 0.28$\pm$0.02} & { ...} & {T10}\\
(38628) Huya & { (6.68/\textbf{6.75}/6.82)$\pm$0.01} & { -} & { $<$0.1} & { 4.7} & { O03b}\\
 &  { -} & { -} & { $<$0.15} & { ...} & { SJ02}\\
 &   { -} & { -} & { $<$0.097} & { ...} & { S02}\\
 &   { -} & { -} & { $<$0.04} & { ...} & { SJ03,LL06}\\
 &   { 5.28} & { -} & { 0.02$\pm$0.01} & { ...} & { TW}\\
(58534) Logos & { -} & { -} & {$\sim$0.8} & { 6.6} & { N08}\\
(90482) Orcus  &  { 7.09/\textbf{10.08$\pm$0.01}/17.43} & { \textbf{20.16}} & { 0.04$\pm$0.02} & { 2.3} & { O06}\\
 &    { 13.19} & { -} & { 0.18$\pm$0.08} & { ...} & { R07}\\
  &   { -} & { -} & { $<$0.03} & { ...} & { S07}\\
  &   { 10.47} & -  & { 0.04$\pm$0.01} & {... } & { T10}\\
(50000) Quaoar   &  { -} & { 17.6788$\pm$0.0004} & { 0.13$\pm$0.03} & { 2.6} & { O03a}\\
&     { 8.84} & { -} & { 0.18$\pm$0.10} & { ...} & { R07}\\
&     { 9.42} & { 18.84} & { $\sim$0.3} & { ...} & { L07}\\
&     { 8.84} & {17.68} & { 0.15$\pm$0.04} & { ...} & { T10}\\
(120347) Salacia   & { -} & { $\sim$17.5} & { 0.2} & { 4.4} & { S10}\\
&    { 6.09 or 8.10} & { -} & { 0.03$\pm$0.01} & { ...} & { T10}\\
&    { 6.5} & { -} & { 0.06$\pm$0.02} & { ...} & { TW}\\
&    {-} & { -} & { $<$0.04} & { ...} & {B13}\\
{(88611B) Sawiskera} &   { 4.7526$\pm$0.0007} & { 9.505$\pm$0.001} & { $\sim$0.6} & { 6.2} & { Os03}\\
&     { 4.749$\pm$0.001} & {9.498$\pm$0.02} & { 0.48$\pm$0.05} & { ...} & { K06b}\\
{(79360) Sila} &  {-} & { -} & { $<$0.08} & { 5.1} & { SJ02}\\
&   { -} & { -} & { $<$0.22} & { ...} & { RT99}\\
&   {150.1194} & { 300.2388} & { 0.14$\pm$0.07} & { ...} & {G12, B13}\\
{(88611) Teharonhiawako}  &   { -} & { -} & { $<$0.15} & { 5.5} & { Os03}\\
    & { 5.50$\pm$0.01} or {7.10$\pm$0.02} & { 11.0$\pm$0.02} or {14.20$\pm$0.04} & (0.32 or 0.30)$\pm$0.04 & { ... } & { K06b}\\
{(42355) Typhon}   &  { (3.66 or 4.35)$\pm$0.02} & { -} & { $<$0.15} & { 7.2} & { O03b}\\
&   { -} & { -} & { $<$0.05} & { ...} & { SJ03}\\
&   { $>$5} & { -} & { -} & { ...} & { D08}\\
&   { 9.67} & - & { 0.07$\pm$0.01} & { ...} & { T10}\\
{(174567) Varda}&   { 5.90 or 7.87} & { -} & { 0.06$\pm$0.01} & { 3.6} & { T10}\\
&    { 5.91} & { -} & { 0.04$\pm$0.01} & { ...} & { TW}\\
&    {-} & { -} & { $<$0.04} & { ...} & {B13}\\
{(26308) 1998~SM$_{165}$} &  { -} & { 7.1$\pm$0.01} & { 0.45$\pm$0.03} & { 5.8} & { SJ02}\\
  &   { 3.983} & {7.966} & { 0.56} & { ...} & { R01}\\
  &   { -} & {8.40$\pm$0.05} & {-} & { ...} & { S06}\\
{(47171) 1999~TC$_{36}$} &  { 6.21$\pm$0.02} & { -} & { 0.06} & { 4.9} & { O03b}\\
&  { -} & { -} & { $<$0.07} & { ...} & { LL06}\\
&   { -} & { -} & { $<$0.05} & { ...} & { SJ03}\\
{(80806) 2000~CM$_{105}$}&   { -} & { -} & { $<$0.14} & { 6.3} & { LL06}\\
{(82075) 2000~YW$_{134}$} &   { -} & { -} & { $<$0.10} & { 5.0} & { SJ03}\\
{2001~QC$_{298}$} &  { -} & { $\sim$12} & { 0.4} & { 6.1} & { S10}\\
  &  {3.89$\pm$0.24} & {7.78$\pm$0.48} & {0.30$\pm$0.04} & {... } & {K06b}\\
{(139775) 2001~QG$_{298}$}&  { 6.8872$\pm$0.0002} & { 13.7744$\pm$0.0004} & { 1.14$\pm$0.04} & { 7.0} & { SJ04}\\
&  { - } & {-  } & { 0.07$\pm$0.01} & { ...} & { L11}\\
{(275809) 2001~QY$_{297}$}  & { 5.84} & {11.68} & { 0.49$\pm$0.03} & { 5.7} & { T12}\\
 & {12.2$\pm$4.3} & { -} & { 0.66$\pm$0.38} & { ... } & {  K06b}\\
{(55637) 2002~UX$_{25}$}& { -} & { 14.382$\pm$0.001 or 16.782$\pm$0.003 } & { 0.21$\pm$0.06} & { 3.6} & { R05b}\\
&  { -} & { -} & { $<$0.06} & { ...} & { SJ03}\\
&  { -} & { -} & { 0.13$\pm$0.09} & { ...} & { R07}\\
{2002~VT$_{130}$}&  { -} & { -} & { $\sim$0.21} & {5.8} & {TW}\\
{(119979) 2002~WC$_{19}$} &  { -} & { -} & { $<$0.05} & { 5.1} & { S07}\\
{2003~AZ$_{84}$}&{ (4.32/5.28/6.72/\textbf{6.76})$\pm$0.01} & { -} & { 0.10$\pm$0.04} & { 3.6} & { O06}\\
& {6.72$\pm$0.05} & { -} & { 0.14$\pm$0.03} & { ...} & { SJ03}\\
&  { 6.79} & { -} & { 0.07$\pm$0.01} & { ...} & { T10}\\
2003~FE$_{128}$ &  { 5.85$\pm$0.15} & { -} & { 0.50$\pm$0.14} & { 6.3} & { K06b}\\
{2003~QY$_{90A}$} & { 3.4$\pm$1.1} & { -} & { 0.34$\pm$0.06} & { 6.3} & { K06a}\\
{2003~QY$_{90B}$}& { 7.1$\pm$2.9} & { -} & { 0.90$\pm$0.18} & { 6.3} & { K06a}\\
{2005~EF$_{298}$} & {4.82} &{ 9.65 }& {0.31$\pm$0.04} & {6.1}& {B13} \\
{(303712) 2005~PR$_{21}$} & {-} &{ - }& {$<$0.28} & {6.1}& {B13} \\
{2007~TY$_{430}$} & - & {9.28} & {0.24$\pm$0.05}   & {6.9}& {TW}\\
{2007~UK$_{126}$} & {11.05} &{ - }& {0.03$\pm$0.01} & {3.4}& {TW} \\

\hline\hline
 
\end{longtable}
\end{onecolumn}
 \end{scriptsize}

References list:\\
B97: \cite{Buie1997}; 
RT99: \cite{RomanishinTegler1999};
SJ02: \cite{SheppardJewitt2002};
O03a: \cite{Ortiz2003Quaoar}; 
O03b: \cite{Ortiz2003};
Os03: \cite{Osip2003};
SJ03: \cite{SheppardJewitt2003};
SJ04: \cite{SheppardJewitt2004}; 
R05b: \cite{Rousselot2005b}; 
K06a: \cite{Kern2006}; 
K06b: \cite{Kern2006_phd};
LL06:\cite{LacerdaLuu2006};
O06:\cite{Ortiz2006}; 
R06: \cite{Rabinowitz2006};
S06: \cite{Spencer2006};
L07: \cite{Lin2007}; 
R07:\cite{Rabinowitz2007}; 
S07: \cite{Sheppard2007};
D08: \cite{Dotto2008}; 
Du08: \cite{Duffard2008};
L08: \cite{Lacerda2008}; 
N08: \cite{Noll2008}; 
R08: \cite{Roe2008}; 
S10: \cite{Snodgrass2010}; 
T10: \cite{Thirouin2010};
L11: \cite{Lacerda2011};
G12: \cite{Grundy2012};
T12: \cite{Thirouin2012};
B13: \cite{Benecchi2013};
TW: This work

 \clearpage

\begin{scriptsize}
\begin{longtable}{lccccccc}  
\caption[Density, size, and albedo from this work and from the literature]{\label{tab:sizealbedo} Density, size, and albedo from this work and from the literature: for each system with short-term variability in this work, in \citet{Thirouin2010}, and \citet{Thirouin2012}, we present the name of the system, the system mass (M$_{syst}$), the lower limit of the density ($\rho$), upper limits of the primary and satellite sizes (R$_{p}$ and R$_{s}$, respectively), and the lower limit of the geometric albedo (p$_{v}$). Several techniques can be used to estimate parameters such as thermal modeling (thermal), determined from mutual orbit of binary components (orbit), direct imaging (direct), occultation, or from the lightcurve. We must point out that when the density is derived from the lightcurve it is only the lower density limit. The last column is dedicated to the references. }\\

 System  &   M$_{syst}$      & $\rho$                   & R$_{p}$    & R$_{s}$       & p$_{v}$  & Technique &  Reference \\
        &  $\times$10$^{18}$[kg]     & [g cm$^{-3}$]        &     [km]       &     [km]      &      & &    \\
\hline
 \hline
\hline
\endfirsthead
 \caption{continued.}\\
 System  &   M$_{syst}$      & $\rho$                   & R$_{p}$    & R$_{s}$       & p$_{v}$   & Technique &  Reference \\
        &  $\times$10$^{18}$[kg]     & [g cm$^{-3}$]        &     [km]       &     [km]      &     & &    \\
\hline
\endhead
\hline
\endfoot

Salacia-Actaea  & 438$\pm$16& 1.16$^{+0.59}_{-0.36}$ & 453$\pm$52 &  152$\pm$18 &  0.0357$^{+0.0103}_{-0.0072}$   & thermal &  S12  \\
			  & ... & 1.38$\pm0.27$ &  431$\pm$35 &  146$\pm$11  &  0.0439$\pm$0.0044   & thermal &   S12, V12  \\
			  & ... & 1.29$^{+0.29}_{-0.23}$& 427$\pm$23  &  143$\pm$12  &  0.044$\pm$0.004   & thermal & S12, F13  \\
			&  ... & >0.92 & <491 & <165 & >0.03     & lightcurve & S12, TW  \\
\hline
Varda-Ilmar\"{e}  &260.2$\pm$4.7 & 1.25$^{+0.40}_{-0.43}$ & 399$\pm$49 &  205$\pm$25 & 0.077$^{+0.025}_{-0.014}$   & thermal & G13, V14  \\
	      		    & ...  & >1.11 & <366 & <188 & >0.09 & lightcurve & G13, TW   \\   
\hline

2007~TY$_{430}$ &  0.790$\pm$0.021  &  0.5 & <60  & <60  & >0.17   & estimate$^{a}$ & Sh12   \\
	      		    & ... & >0.46 &<58& <55 & >0.12 & lightcurve & Sh12, TW   \\   
\hline

2001~QY$_{297}$ & 4.10$\pm$0.04  &  0.92$^{+1.30}_{-0.27}$ & 85$^{+8}_{-40}$ & 77$^{+8}_{-37}$ & 0.075$^{+0.037}_{-0.027}$   & thermal & G11b, V12, V14   \\
			&  ... & >0.29 & <129 & <107& >0.08     & lightcurve & G11b, T12   \\
\hline

Huya &  ? &   ? &  236$\pm$11 & 124$\pm$6 & 0.05$\pm$0.05   & thermal & S08   \\
  & ? &   ? &  195$\pm$12 & 102$\pm$6  & 0.081$\pm$0.011   & thermal & M12   \\
  & ? &   >1.43 &  ? &? & ?  & lightcurve & TW   \\

\hline
Orcus-Vanth  &  632$\pm$1&  1.5$\pm$0.3 & 450$\pm$34   &  136$\pm$10  & 0.28$\pm$0.04   & thermal+orbit  & B10   \\
	               & ... &   ? & 453$\pm$36 &  137$\pm$11 & 0.197$\pm$0.034   & thermal & B10, S08   \\
	               & ... &   ? & 416.5$\pm$22.5 & 126$\pm$8 & 0.25$\pm$0.03   & thermal & B10, L10   \\
			  & ... & 1.53$^{+0.15}_{-0.13}$& 459$\pm$13 & 138$\pm$9  & 0.231$^{+0.018}_{-0.011}$   & thermal & B10, F13   \\
			  &  ... & >0.35 & <749 & <225 & >0.09    & lightcurve & B10, T10, TW   \\
\hline

Typhon-Echidna  &  0.949$\pm$0.052 & 0.44$^{+0.44}_{-0.17}$   &  78$\pm$8 & 43$\pm$4 &  ? & thermal+orbit &   G08  \\
	      		    &... & ? & 76$\pm$8 & 42$\pm$4 & 0.051$^{+0.012}_{-0.008}$& thermal & G08, S08   \\  
	      		    &... & ? & 61$\pm$4 & 34$\pm$3 & 0.08$\pm$0.01  & thermal & G08, M10 \\  
	      		    & ... & 0.36$^{+0.08}_{-0.07}$ & 81$\pm$4 & 45$\pm$3 & 0.044$\pm$0.003 & thermal & G08, SS12   \\   
	      		    &... & >0.42 & <76 & <42 & >0.06 & lightcurve & G08, T10, TW   \\   
 \hline

Quaoar-Weywot  &  ?  &  ?  &  633$\pm$100 & 47$\pm$7 &  0.092$^{+0.036}_{-0.023}$  & direct &  B04  \\
                         & ?         &?    &   425$\pm$99 &  33$\pm$8    & 0.199$^{+0.132}_{-0.070}$     & thermal & S08   \\	         
                         & ?         &?    &   450$\pm$56 &  34$\pm$4    & 0.172$^{+0.055}_{-0.036}$     &   thermal & B09   \\	  
			     & 1600$\pm$300     & 4.2$\pm$1.3 &  445$\pm$35 &  34$\pm$3  & ?  & orbit & F10  \\
			     & (1300-1500)$\pm$100     & 2.7-5.0 &  ?  &  ?  & ?  & orbit  & Fr13  \\
			     & 1650$\pm$160     & 1.6$\pm$1.3 or 4.5$\pm$1.8 & ?   &  ?  &   ? & orbit$^{b}$  & VBM12  \\
			     & (1300-1500)$\pm$100     & 2.18$^{+0.43}_{-0.36}$  &  535$\pm$19  &  41$\pm$6  & 0.137$^{+0.011}_{-0.013}$  & thermal  & Fr13, F13  \\
			     & (1300-1500)$\pm$100     & 1.99$\pm$0.46  &  569$^{+24}_{-17}$  &  41$\pm$6  & 0.109$\pm$0.007  & occultation$^{c}$  & Fr13, B13  \\
			     & 1600$\pm$300                & >0.5 & <914 & <69 &>0.06     & lightcurve & F10, T10, TW   \\

\hline
2003~AZ$_{84}$  &  ?   & ? &  446$\pm$28  &  45$\pm$3 &   0.065$\pm$0.008  & thermal &  M10 \\
				&  ?   & ? &  362$\pm$32  &  36$\pm$15 &   0.107$^{+0.023}_{-0.016}$  & thermal &  M12 \\
 	      		    & ? & >0.85 & ? & ? & ? & lightcurve & T10   \\   
\hline
2007~UK$_{126}$  &  ?   & ?  &  295$\pm$38  & 52$\pm$7  &    0.167$^{+0.058}_{-0.038}$   & thermal  & SS12  \\
	      		    & ? & >0.32 & ? & ? & ? & lightcurve & TW   \\   
 

\hline

\hline

\end{longtable}
 \end{scriptsize}
Notes: \\
$^{a}$: \citet{Sheppard2012} only assumed a possible density to derive a size and albedo, none of these values were computed.\\
$^{b}$: Two orbital solutions were suggested by \citet{Vachier2012}, and two different density estimates were derived depending on the orbital solution. \\
$^{c}$: \citet{BragaRibas2013} obtained the equatorial radius of Quaoar. \\

References list:
B04: \citet{Brown2004}; 
G08: \citet{Grundy2008}; 
S08: \citet{Stansberry2008}; 
B09: \citet{Brucker2009}; 
B10: \citet{Brown2010}; 
F10: \citet{Fraser2010}; 
L10: \citet{Lim2010}; 
M10: \citet{Muller2010}; 
T10: \citet{Thirouin2010};
G11: \citet{Grundy2011DPS};
G11b: \citet{Grundy2011}; 
M12: \citet{Mommert2012}; 
S12: \citet{Stansberry2012}; 
SS12: \citet{SantosSanz2012}; 
Sh12: \citet{Sheppard2012}; 
T12: \citet{Thirouin2012};
V12: \citet{Vilenius2012}; 
VBM12: \citet{Vachier2012};
B13: \citet{BragaRibas2013};
F13: \citet{Fornasier2013};
Fr13: \citet{Fraser2013};
G13: \citet{Grundy2013};
V14: \citet{Vilenius2014};
TW: This work.

\clearpage

 \begin{scriptsize}
 
\begin{longtable}{lccccccccccc}

\caption{\label{Tab:tidallockgladman} System names, Love number of the primary (\textit{k}$_{primary}$), primary and satellite densities ($\rho$$_{primary}$ and $\rho$$_{satellite}$), initial rotational rate of the primary (\textit{T}$_{0}$), mean orbital angular velocity (\textit{n}), and orbital semimajor axis (\textit{a}). We assume a density of 0.5-1~g cm$^{-3}$ for the satellite, except in some cases (see discussion section). The initial rotational period of the primary is the breakup rotation rate (upper limit) expressed as: T$_{0}$=(3$\pi$/G$\rho$$_{primary}$)$^{1/2}$. Equation~\ref{Eq:Circular} was used to compute the time needed to circularize the orbit ($\tau _{circular}$). Equation~\ref{Eq:LockGladman} was used to compute the time needed to tidally lock the primary ($\tau _{lock}$).}\\

\hline 
System & g$_{primary}$ &$\mu$ &  k$_{primary}$& $\rho$$_{primary}$ & $\rho$$_{satellite}$ &  T$_{0}$ & n  &  a & Q & $\tau _{lock}$ & $\tau _{circular}$\\
    & [m s$^{-2}$] & [N m$^{-2}$] &  & [g cm$^{-3}$]  & [g cm$^{-3}$] &  [h] & [rad s$^{-1}$] & [km] & & [years] &  [years] \\
\hline 
\hline 
Salacia- &   &  & & & &  &  & & &  &  \\
Actaea &   &  & & & &  &  & & &  &  \\
 & (1.3-2)$\times$10$^{-1}$ & 4$\times$10$^{9}$ & (2.5-5.5)$\times$10$^{-3}$ & 1.38$\pm$0.27 &  1.38$\pm$0.27& 2.57-3.13 & (1-1.7)$\times$10$^{-5}$ & 5,619$\pm$87 &100 & (0.4-1)$\times$10$^{9}$ & (0.4-2.1)$\times$10$^{8}$  \\
 &  (1.3-2)$\times$10$^{-1}$& 3$\times$10$^{10}$ &  (3.3-7.3)$\times$10$^{-4}$ & 1.38$\pm$0.27 &  1.38$\pm$0.27 & 2.57-3.13& (1-1.7)$\times$10$^{-5}$ &5,619$\pm$87&100 & (3-9)$\times$10$^{9}$  & (0.3-1.6)$\times$10$^{9}$ \\
 &(1.3-2)$\times$10$^{-1}$ & 4$\times$10$^{9}$ & (2.5-5.5)$\times$10$^{-3}$ & 1.38$\pm$0.27 &  1.00 & 2.57-3.13 & (1-1.7)$\times$10$^{-5}$  &5,619$\pm$87&100 & (1.2-1.5)$\times$10$^{9}$  & (0.7-2.3)$\times$10$^{8}$ \\
  & (1.3-2)$\times$10$^{-1}$& 4$\times$10$^{9}$ & (2.5-5.5)$\times$10$^{-3}$& 1.38$\pm$0.27 &  0.50 & 2.57-3.13 & (1-1.7)$\times$10$^{-5}$  &5,619$\pm$87&100 & (5-6)$\times$10$^{9}$  & (1.3-4.6)$\times$10$^{8}$ \\
  &  (1.3-2)$\times$10$^{-1}$& 3$\times$10$^{10}$ &  (3.3-7.3)$\times$10$^{-4}$ & 1.38$\pm$0.27 & 1.00 & 2.57-3.13 & (1-1.7)$\times$10$^{-5}$ &5,619$\pm$87&100 & (0.9-1.1)$\times$10$^{10}$  & (0.5-1.7)$\times$10$^{9}$ \\ 
 &  (1.3-2)$\times$10$^{-1}$ & 3$\times$10$^{10}$ & (3.3-7.3)$\times$10$^{-4}$ & 1.38$\pm$0.27 & 0.50 & 2.57-3.13 & (1-1.7)$\times$10$^{-5}$  &5,619$\pm$87&100 & (3.7-4.6)$\times$10$^{10}$  & (1-1.7)$\times$10$^{9}$  \\

 \hline
Huya & (7.3-8.3)$\times$10$^{-2}$ & 4$\times$10$^{9}$ & (7.6-9.7)$\times$10$^{-4}$ & 1.43 &  1.43 & 2.76 &   (2-2.5)$\times$10$^{-5}$ &  1,800 &100 & (1.9-3)$\times$10$^{7}$   & (0.9-1.3)$\times$10$^{7}$ \\
Huya &  (7.3-8.3)$\times$10$^{-2}$  & 3$\times$10$^{10}$ & (1-1.3)$\times$10$^{-4}$ & 1.43 &  1.43 & 2.76 &  (2-2.5)$\times$10$^{-5}$ & 1,800 & 100 & (1.4-2.3)$\times$10$^{8}$   & (7-9.7)$\times$10$^{7}$  \\
Huya &  (7.3-8.3)$\times$10$^{-2}$  & 4$\times$10$^{9}$ & (7.6-9.7)$\times$10$^{-4}$ & 1.43 &  1.00 & 2.76 & (2-2.5)$\times$10$^{-5}$ &1,800&100 & (3.9-6.1)$\times$10$^{7}$  & (1.3-1.8)$\times$10$^{7}$\\
Huya & (7.3-8.3)$\times$10$^{-2}$  & 4$\times$10$^{9}$ & (7.6-9.7)$\times$10$^{-4}$ & 1.43 &  0.50 & 2.76 & (2-2.5)$\times$10$^{-5}$ &1,800&100 & (0.6-1.6)$\times$10$^{8}$   & (2.7-3.7)$\times$10$^{7}$\  \\
Huya  & (7.3-8.3)$\times$10$^{-2}$  & 3$\times$10$^{10}$ & (1-1.3)$\times$10$^{-4}$ & 1.43 &  1.00 & 2.76 & (2-2.5)$\times$10$^{-5}$ &1,800&100 & (2.9-4.6)$\times$10$^{8}$  & (1-1.4)$\times$10$^{8}$ \\
Huya & (7.3-8.3)$\times$10$^{-2}$  & 3$\times$10$^{10}$ & (1-1.3)$\times$10$^{-4}$ & 1.43 &  0.50 & 2.76 &(2-2.5)$\times$10$^{-5}$ &1,800 &100 & (1.2-1.8)$\times$10$^{9}$  & (2-2.8)$\times$10$^{8}$ \\
 
 \hline
Varda- &   &  & & & &  &  & & &  &  \\
Ilmar\"{e}  &   &  & & & &  &  & & &  &  \\
  &  (1.1-1.4)$\times$10$^{-1}$ & 4$\times$10$^{9}$ & (1.6-2.7)$\times$10$^{-3}$ & 1.11 &  1.11 & 3.13 & (1.3-1.9)$\times$10$^{-5}$  &4,200 &100 & (1.7-4.6)$\times$10$^{7}$  & (1-1.8)$\times$10$^{7}$  \\
   & (1.1-1.4)$\times$10$^{-1}$  & 3$\times$10$^{10}$ & (2.2-3.6)$\times$10$^{-4}$ & 1.11 &  1.11 & 3.13 & (1.3-1.9)$\times$10$^{-5}$ &4,200& 100 & (1.3-3.4)$\times$10$^{8}$ & (0.7-1.3)$\times$10$^{8}$ \\
 &(1.1-1.4)$\times$10$^{-1}$ & 4$\times$10$^{9}$ & (1.6-2.7)$\times$10$^{-3}$ & 1.11 &  1.00 & 3.13 & (1.3-1.9)$\times$10$^{-5}$  & 4,200&100 & (2.1-5.6)$\times$10$^{7}$  & (1.1-2)$\times$10$^{7}$  \\
 &(1.1-1.4)$\times$10$^{-1}$  & 4$\times$10$^{9}$ & (1.6-2.7)$\times$10$^{-3}$ & 1.11 &  0.50 & 3.13 & (1.3-1.9)$\times$10$^{-5}$ &4,200 &100 & (0.8-2.2)$\times$10$^{8}$  & (2.1-3.9)$\times$10$^{7}$  \\
  &  (1.1-1.4)$\times$10$^{-1}$ & 3$\times$10$^{10}$ & (2.2-3.6)$\times$10$^{-4}$ & 1.11 &  1.00 & 3.13 & (1.3-1.9)$\times$10$^{-5}$ &4,200& 100 & (1.6-4.2)$\times$10$^{8}$  & (0.8-1.5)$\times$10$^{8}$ \\
  & (1.1-1.4)$\times$10$^{-1}$  & 3$\times$10$^{10}$ & (2.2-3.6)$\times$10$^{-4}$ & 1.11 &  0.50 & 3.13 & (1.3-1.9)$\times$10$^{-5}$ &4,200& 100 & (0.6-1.7)$\times$10$^{9}$  &(1.6-2.9)$\times$10$^{8}$  \\
\hline

2007~UK$_{126}$ & (2.3-3)$\times$10$^{-2}$ & 4$\times$10$^{9}$ & (7.5-12.5)$\times$10$^{-5}$ & 0.32 &  0.32  & 5.84 & (5.7-8.4)$\times$10$^{-6}$  & 3,600 & 100 & (3-9.8)$\times$10$^{12}$ & (2.5-4.7)$\times$10$^{10}$  \\
2007~UK$_{126}$ &  (2.3-3)$\times$10$^{-2}$ & 3$\times$10$^{10}$ & (0.9-1.7)$\times$10$^{-5}$ &0.32  &  0.32  & 5.84 & (5.7-8.4)$\times$10$^{-6}$ &3,600&100 & (0.7-2.2)$\times$10$^{13}$ & (1.9-3.5)$\times$10$^{11}$ \\
\hline

2007~TY$_{430}$ &    7.5$\times$10$^{-3}$ & 4$\times$10$^{9}$ & 7.9$\times$10$^{-6}$ & 0.46 &  0.46 & 4.87 & (5.1-5.3)$\times$10$^{-8}$  &21,000$\pm$160&100 & (4.4-4.9)$\times$10$^{17}$  &  (7.6-8.3)$\times$10$^{18}$ \\
2007~TY$_{430}$ &   7.5$\times$10$^{-3}$ & 3$\times$10$^{10}$ & 1.05$\times$10$^{-6}$ & 0.46 &  0.46 & 4.87 &  (5.1-5.3)$\times$10$^{-8}$ & 21,000$\pm$160&100 & (3.3-3.6)$\times$10$^{18}$  & (5.7-6.3)$\times$10$^{19}$ \\
\hline

Typhon- &   &  & & & &  &  & & &  &  \\
Echidna &   &  & & & &  &  & & &  &  \\
 &   (0.6-1.1)$\times$10$^{-2}$  & 4$\times$10$^{9}$& (0.6-1.5)$\times$10$^{-5}$& 0.36$^{+0.08}_{-0.07}$  & 0.36$^{+0.08}_{-0.07}$ &  4.98-6.13 & (2.9-4.3)$\times$10$^{-6}$  & 1,628$\pm$29 &100 & (2.4-4.6)$\times$10$^{11}$ & (1.3-5.4)$\times$10$^{11}$  \\
 &   (0.6-1.1)$\times$10$^{-2}$   & 3$\times$10$^{10}$& (0.7-2.1)$\times$10$^{-6}$& 0.36$^{+0.08}_{-0.07}$  & 0.36$^{+0.08}_{-0.07}$ &  4.98-6.13 & (2.9-4.3)$\times$10$^{-6}$   &1,628$\pm$29& 100 &  (1.8-3.4)$\times$10$^{12}$  &  (1-4)$\times$10$^{12}$  \\
\hline
Quaoar- &   &  & & & &  &  & & &  &  \\
Weywot &   &  & & & &  &  & & &  &  \\
  & (3.6-4.9)$\times$10$^{-2}$ & 4$\times$10$^{9}$&  (1.8-3.4)$\times$10$^{-4}$ &  2.7-5.0 & 2.7-5.0   &  1.48-2.01 &  1.5$\times$10$^{-7}$ &  14,500$\pm$800 &100  & (0.2-1.2)$\times$10$^{9}$ &  (1.6-3.8)$\times$10$^{13}$  \\
  & (3.6-4.9)$\times$10$^{-2}$  & 3$\times$10$^{10}$&  (2.4-4.5)$\times$10$^{-5}$ &  2.7-5.0 & 2.7-5.0   &  1.48-2.01 &  1.5$\times$10$^{-7}$  &14,500$\pm$800& 100  & (1.2-9.3)$\times$10$^{9}$& (1.2-2.9)$\times$10$^{14}$ \\
  & (3.6-4.9)$\times$10$^{-2}$ & 4$\times$10$^{9}$ &  (1.8-3.4)$\times$10$^{-4}$ &  2.7-5.0 & 1.00   &  1.48-2.01 &  1.5$\times$10$^{-7}$  &14,500$\pm$800&100  & (4-9)$\times$10$^{9}$ &(0.8-1)$\times$10$^{14}$\\
 & (3.6-4.9)$\times$10$^{-2}$  & 3$\times$10$^{10}$&  (2.4-4.5)$\times$10$^{-5}$ &  2.7-5.0 & 1.00   &  1.48-2.01 & 1.5$\times$10$^{-7}$ &14,500$\pm$800&100  & (3-6.8)$\times$10$^{10}$ &  (6.1-7.7)$\times$10$^{14}$\\
 & (3.6-4.9)$\times$10$^{-2}$ & 4$\times$10$^{9}$&  (1.8-3.4)$\times$10$^{-4}$ &  2.7-5.0 & 0.50   &  1.48-2.01 &  1.5$\times$10$^{-7}$ &14,500$\pm$800&100  & (1.6-3.6)$\times$10$^{10}$  & (1.6-2.1)$\times$10$^{14}$ \\
 & (3.6-4.9)$\times$10$^{-2}$  & 3$\times$10$^{10}$&  (2.4-4.5)$\times$10$^{-5}$ &  2.7-5.0 & 0.50  &  1.48-2.01 &1.5$\times$10$^{-7}$ & 14,500$\pm$800&100  & (1.2-2.7)$\times$10$^{11}$ &(1.2-1.5)$\times$10$^{15}$\\
 & (2-2.4)$\times$10$^{-2}$ & 4$\times$10$^{9}$&  (5.7-8.1)$\times$10$^{-5}$ &  1.99$\pm0.46$ &1.99$\pm0.46$  &  2.11-2.67 & 1.1$\times$10$^{-7}$ &14,500$\pm$800&100  & (1-5)$\times$10$^{9}$&  (1-1.6)$\times$10$^{14}$ \\
 & (2-2.4)$\times$10$^{-2}$  & 3$\times$10$^{10}$& (0.8-1.1)$\times$10$^{-5}$ &  1.99$\pm0.46$ &1.99$\pm0.46$  & 2.11-2.67  &  1.1$\times$10$^{-7}$ & 14,500$\pm$800&100  & (0.7-3.9)$\times$10$^{10}$ & (0.7-1.2)$\times$10$^{15}$ \\
 & (2-2.4)$\times$10$^{-2}$ & 4$\times$10$^{9}$&  (5.7-8.1)$\times$10$^{-5}$ &  1.99$\pm0.46$ &1.00 &  2.11-2.67 & 1.1$\times$10$^{-7}$ &14,500$\pm$800&100  & (0.6-1.2)$\times$10$^{10}$ & (2.3-2.4)$\times$10$^{14}$  \\
 & (2-2.4)$\times$10$^{-2}$  & 3$\times$10$^{10}$&  (0.8-1.1)$\times$10$^{-5}$ &  1.99$\pm0.46$ &1.00 & 2.11-2.67  & 1.1$\times$10$^{-7}$ &14,500$\pm$800&100  & (4.3-9)$\times$10$^{10}$ &  (1.7-1.8)$\times$10$^{15}$\\ 
 & (2-2.4)$\times$10$^{-2}$ & 4$\times$10$^{9}$&  (5.7-8.1)$\times$10$^{-5}$ &  1.99$\pm0.46$ &0.50 &  2.11-2.67 & 1.1$\times$10$^{-7}$ &14,500$\pm$800&100  & (2.3-4.8)$\times$10$^{10}$ &  (4.7-4.8)$\times$10$^{14}$\\
 & (2-2.4)$\times$10$^{-2}$  & 3$\times$10$^{10}$& (0.8-1.1)$\times$10$^{-5}$ &  1.99$\pm0.46$ &0.50 & 2.11-2.67  &1.1$\times$10$^{-7}$  &14,500$\pm$800&100  & (1.7-3.6)$\times$10$^{11}$ &(3.5-3.6)$\times$10$^{15}$ \\
  & (2.3-2.6)$\times$10$^{-2}$ & 4$\times$10$^{9}$&  (7.2-9.2)$\times$10$^{-5}$ &  2.18$^{+0.43}_{-0.36}$ & 2.18$^{+0.43}_{-0.36}$&  2.04-2.45 &  (1.1-1.2)$\times$10$^{-7}$ &14,500$\pm$800 &100  & (1.2-5.7)$\times$10$^{9}$&  (1-1.4)$\times$10$^{14}$\\
  & (2.3-2.6)$\times$10$^{-2}$  & 3$\times$10$^{10}$& (1-1.2)$\times$10$^{-5}$ &  2.18$^{+0.43}_{-0.36}$ &2.18$^{+0.43}_{-0.36}$ & 2.04-2.45  &  (1.1-1.2)$\times$10$^{-7}$ &14,500$\pm$800&100  & (0.9-4.3)$\times$10$^{10}$& (0.8-1.1)$\times$10$^{15}$\\
  & (2.3-2.6)$\times$10$^{-2}$ & 4$\times$10$^{9}$&  (7.2-9.2)$\times$10$^{-5}$ &  2.18$^{+0.43}_{-0.36}$ & 1.00 &  2.04-2.45 &  (1.1-1.2)$\times$10$^{-7}$  &14,500$\pm$800& 100  & (0.8-1.7)$\times$10$^{10}$ & (2.5-2.6)$\times$10$^{14}$   \\
  & (2.3-2.6)$\times$10$^{-2}$  & 3$\times$10$^{10}$& (1-1.2)$\times$10$^{-5}$ &  2.18$^{+0.43}_{-0.36}$ & 1.00 & 2.04-2.45  &(1.1-1.2)$\times$10$^{-7}$ &14,500$\pm$800&100  & (0.6-1.3)$\times$10$^{11}$& (1.9-2)$\times$10$^{15}$\\
 & (2.3-2.6)$\times$10$^{-2}$ & 4$\times$10$^{9}$&  (7.2-9.2)$\times$10$^{-5}$ &  2.18$^{+0.43}_{-0.36}$ & 0.50 &  2.04-2.45 &(1.1-1.2)$\times$10$^{-7}$ &14,500$\pm$800&100  & (3.3-6.8)$\times$10$^{10}$ &(5-5.3)$\times$10$^{14}$\\
  & (2.3-2.6)$\times$10$^{-2}$  & 3$\times$10$^{10}$&(1-1.2)$\times$10$^{-5}$  &  2.18$^{+0.43}_{-0.36}$ & 0.50 & 2.04-2.45  &(1.1-1.2)$\times$10$^{-7}$ &14,500$\pm$800&100  & (2.5-5.1)$\times$10$^{11}$ &  (3.7-4)$\times$10$^{15}$\\

\hline

Orcus- &   &  & & & &  &  & & &  &  \\
Vanth &   &  & & & &  &  & & &  &  \\
  & (5.8-6.1)$\times$10$^{-2}$ & 4$\times$10$^{9}$&  (0.5-5.2)$\times$10$^{-3}$ &  1.53$^{+0.15}_{-0.13}$ & 1.53$^{+0.15}_{-0.13}$  &  2.55-2.79 & (1.2-1.4)$\times$10$^{-6}$ &8,980$\pm$23&100  & (5.7-9.8)$\times$10$^{7}$&  (2.2-2.4)$\times$10$^{10}$ \\
 &   (5.8-6.1)$\times$10$^{-2}$ & 3$\times$10$^{10}$&  (6.2-6.9)$\times$10$^{-5}$ &  1.53$^{+0.15}_{-0.13}$  & 1.53$^{+0.15}_{-0.13}$   &  2.55-2.79  &(1.2-1.4)$\times$10$^{-6}$&8,980$\pm$23& 100  & (4.2-7.3)$\times$10$^{8}$ &  (1.7-1.8)$\times$10$^{11}$ \\
   & (5.8-6.1)$\times$10$^{-2}$  & 4$\times$10$^{9}$&  (3.6-4.7)$\times$10$^{-4}$ &  1.53$^{+0.15}_{-0.13}$ & 1.00  &  2.55-2.79 & (1.2-1.4)$\times$10$^{-6}$&8,980$\pm$23& 100  & (1.7-1.9)$\times$10$^{8}$ & (3.1-5.3)$\times$10$^{10}$ \\
 &  (5.8-6.1)$\times$10$^{-2}$  & 3$\times$10$^{10}$&  (4.8-6.2)$\times$10$^{-5}$ &  1.53$^{+0.15}_{-0.13}$  & 1.00  &  2.55-2.79 & (1.2-1.4)$\times$10$^{-6}$&8,980$\pm$23& 100  & (1.3-1.4)$\times$10$^{9}$ & (2.4-4)$\times$10$^{11}$ \\ 
&  (5.8-6.1)$\times$10$^{-2}$  & 4$\times$10$^{9}$&  (4.7-5.2)$\times$10$^{-3}$ &  1.53$^{+0.15}_{-0.13}$ & 0.50  &  2.55-2.79 & (1.2-1.4)$\times$10$^{-6}$&8,980$\pm$23& 100  & (6.3-7.7)$\times$10$^{8}$&  (6.3-8)$\times$10$^{10}$ \\
 &   (5.8-6.1)$\times$10$^{-2}$  & 3$\times$10$^{10}$&  (6.2-6.9)$\times$10$^{-5}$ &  1.53$^{+0.15}_{-0.13}$  & 0.50  &  2.55-2.79 & (1.2-1.4)$\times$10$^{-6}$&8,980$\pm$23& 100  & (4.7-5.7)$\times$10$^{9}$ & (4.7-6)$\times$10$^{11}$ \\
\hline

2001~QY$_{297}$ &  (4-7.6)$\times$10$^{-3}$ & 4$\times$10$^{9}$ &  (1.9-8)$\times$10$^{-6}$ & 0.29 & 0.29 &  6.13 & (0.9-2.6)$\times$10$^{-7}$ & 9,960$\pm$31 & 100  &  (0.2-4.3)$\times$10$^{16}$ & (0.4-2.2)$\times$10$^{17}$ \\
2001~QY$_{297}$ & (4-7.6)$\times$10$^{-3}$  & 3$\times$10$^{10}$&  (0.3-1.1)$\times$10$^{-6}$& 0.29  & 0.29 & 6.13 & (0.9-2.6)$\times$10$^{-7}$  & 9,960$\pm$31& 100  & (0.2-3.2)$\times$10$^{17}$ & (0.3-1.7)$\times$10$^{18}$ \\
\hline

2003~AZ$_{84}$ &  (0.5-1)$\times$10$^{-2}$ & 4$\times$10$^{9}$ &  (0.4-2.1)$\times$10$^{-5}$ & 0.85  & 0.85 & 3.58  & (0.8-2.9)$\times$10$^{-7}$  & 7,200  &100  & (4.5-9.2)$\times$10$^{9}$&  (0.02-1.2)$\times$10$^{15}$\\
2003~AZ$_{84}$ &  (0.5-1)$\times$10$^{-2}$ & 3$\times$10$^{10}$ &  (0.5-2.8)$\times$10$^{-6}$  & 0.85  & 0.85  & 3.58   & (0.8-2.9)$\times$10$^{-7}$  & 7,200&100 & (3.4-6.9)$\times$10$^{10}$  &   (0.1-8.9)$\times$10$^{15}$  \\
2003~AZ$_{84}$ & (0.5-1)$\times$10$^{-2}$ & 4$\times$10$^{9}$ &  (0.3-2.1)$\times$10$^{-5}$ & 0.85  & 0.50 & 3.58  & (0.8-2.9)$\times$10$^{-7}$  & 7,200&100  &  (1.3-2.7)$\times$10$^{10}$ & (0.03-2)$\times$10$^{15}$ \\
2003~AZ$_{84}$ & (0.5-1)$\times$10$^{-2}$& 3$\times$10$^{10}$ &   (0.5-2.8)$\times$10$^{-6}$  & 0.85  & 0.50  & 3.58   & (0.8-2.9)$\times$10$^{-7}$  & 7,200&100 & (1-2)$\times$10$^{11}$&  (0.02-1.5)$\times$10$^{16}$ \\

\hline
\hline
\end{longtable}
Notes: \\
Orbital semimajor axis are taken from 
\citet{WeyvotMPC}, \citet{Grundy2008}, \citet{Ragozzine2009}, \citet{Brown2010}, \citet{Fraser2010}, \citet{Grundy2011DPS}, \citet{Grundy2011}, \citet{Ortiz2011}, \citet{Noll2012}, \citet{Sheppard2012}, and \citet{Stansberry2012}.
  \end{scriptsize}

 \begin{scriptsize}

\begin{longtable}{lcccccccccccccccccccc} 
\caption{\label{Tab:samssr} System names, secondary-to-primary mass ratio (q), semi-major axis (a), eccentricity (e), primary radius (R$_{p}$), rotational period of the primary and its critical rotational period (P$_{p}$, and P$_{c}$), and the $\lambda$ shape parameter of the primary ($\lambda$$_{p}$) and satellite. Here, we consider the satellites as spherical bodies, therefore $\lambda$$_{s}$=1. The specific angular momentum (H) and the scaled spin rate (SSR) are also reported.}\\
\hline

System &  q & a &  e & R$_{p}$ & $\lambda$$_{p}$     &  P$_{p}$ & P$_{c}$       &   H&      SSR   & Ref. \\
             &    &  [km]  &    & [km]   &        & [h]   &    [h]   &              &  &    \\
\hline
\hline
\endfirsthead
\caption{continued.}\\
System &  q & a &  e & R$_{p}$ & $\lambda$$_{primary}$     &  P$_{primary}$ & P$_{critical}$           & H&      SSR       & Ref.   \\
             &     &  [km]  &    & [km]   &        & [h]   &    [h]   &              &  &    \\
\endhead
\hline
\endfoot
Haumea-Namaka &   0.001  &   25,657$\pm$91 &  0.249$\pm$0.015   & 709$\pm$50 &    1.49     & 3.92 & 2.09 & 0.32$\pm$0.05 & 0.61$\pm$0.05  & R09, T10\\

\hline
Haumea-Hi'iaka&     0.01    &    49,880$\pm$198   &  0.0513$\pm$0.0078   &  709$\pm$50 &  1.49     & 3.92  &  2.09 & 0.40$\pm$0.05  & 0.61$\pm$0.05      &R09, T10\\   

\hline
 
Orcus-Vanth  &   0.03   &   8,980$\pm$23  &  0.001$\pm$0.001   &  459$\pm$13 &  1.44      & 10.47   & 2.67$\pm$0.12 & 0.26$\pm$0.02 & 0.29$\pm$0.06 & B10, T10\\   
Orcus-Vanth  &   0.09   &   8,980$\pm$23  &  0.001$\pm$0.001   & 459$\pm$13   &   1.44    & 10.47   &  2.67$\pm$0.12 & 0.46$\pm$0.02 & 0.29$\pm$0.06  & O11, T10\\   

\hline

Salacia-Actaea  & 0.04    &  5,619$\pm$87   &   0.02$\pm$0.04  & 427$\pm$23  &  1.44   &  6.5   &  2.81$\pm$0.56  &   0.36$\pm$0.07 & 0.49$\pm$0.14 & S12, T12\\   

\hline

Varda-Ilmar\"{e} $^{a}$    &   0.14   &  4,200   &   0.0084   & 399$\pm$46  &  1.43      & 5.91 & 3.13  &  0.59  & 0.61  &G11b, T13, T12\\   

\hline

2007~TY$_{430}$ $^{b}$ &   >0.85     &  21,000$\pm$160   & 0.1529$\pm$0.0028    & <58 &  1.47       &  9.28  &  4.87    & 4.33 &  0.61  & Sh12, T12\\  

\hline 

Quaoar-Weywot &  0.0004  &    14,500$\pm$800 &  $\sim$0.13-0.16   & 569$^{+24}_{-17}$ &  1.45   &  8.84  & 2.39$\pm$0.28 &   0.15$\pm$0.02    &   0.31$\pm$0.12 & F10, T12\\   

\hline 

Typhon-Echidna  &  0.17  &  1,628$\pm$29   &   0.53$\pm$0.01  &  76$^{+14}_{-16}$ &  1.44 & 9.67  & 5.60$\pm$0.58 & 0.73$\pm$0.06   & 0.66$\pm$0.16 &  G08, T12\\   

\hline 

2001~QY$_{297}$ &  0.56   &  9,960$\pm$31   &  0.418$\pm$0.002   &  130$\pm$21 &  1.62     &  11.68  & 6.13$\pm$1.5 & 1.85$\pm$0.39 & 0.58$\pm$0.21  & G11, T12\\  

\hline
\hline
\end{longtable} 
 \end{scriptsize}
Notes: \\
$^{a}$: The orbit of Ilmar\"{e} is unknown, and only values from the discovery circular are used here. The specific angular momentum and scaled spin rate were computed using a lower limit to the density of 1.11~g cm$^{-3}$.   \\ 
$^{b}$: The specific angular momentum and scaled spin rate were computed using a lower limit to the density of 0.46~g cm$^{-3}$, and upper limits to the component sizes.   \\    
Reference list:
G08: \citet{Grundy2008};
R09: \citet{Ragozzine2009}; 
B10: \citet{Brown2010}; 
F10: \citet{Fraser2010};
T10: \citet{Thirouin2010}; 
G11: \citet{Grundy2011};
G11b: \citet{Grundy2011DPS};
S12: \citet{Stansberry2012};
T12: \citet{Thirouin2012};
Sh12: \citet{Sheppard2012};


\end{document}